\documentclass[11pt,reqno]{amsart}
\usepackage{newtxtext} 

\usepackage[foot]{amsaddr}
\usepackage{amsaddr}
\usepackage{mathptmx}

\usepackage[T1]{fontenc}
\usepackage{amsmath,amscd}

\usepackage[colorlinks=true, pdfstartview=FitV, linkcolor=blue,
            citecolor=blue, urlcolor=blue]{hyperref}  

\usepackage{amsmath,amscd}
\usepackage{bm,epsfig,ifthen}
\usepackage{amsfonts,amssymb}
\usepackage{colortbl}
\usepackage{pstricks}
\usepackage{mathtools}
\usepackage{algorithm,algorithmic}

\usepackage[font=small,labelfont=bf]{caption}
\usepackage{color}
\usepackage{graphicx,graphics}
\usepackage{amsfonts}

\usepackage{needspace} 

\definecolor{rred}{rgb}{0.7,0,0.1}
\definecolor{ccyan}{rgb}{0,.5,1}
\definecolor{ccyan}{rgb}{0,.5,1}
\definecolor{greenrb}{rgb}{0.2,0.6,0.2}

\date{}

\makeatletter
\renewcommand{\fnum@figure}{\textbf{Figure \thefigure}}
\renewcommand{\fnum@table}{\textbf{Table \thetable}}
\makeatother
\newtheorem{proposition}{Proposition}

\def\bi{\begin{itemize}}
\def\ei{\end{itemize}}
\def\be{\begin{equation}}
\def\ee{\end{equation}}
\def\bea{\begin{equation} \begin{aligned}}
\def\eea{\end{aligned} \end{equation}}
\def\beas{\begin{equation*} \begin{aligned}}
\def\eeas{\end{aligned} \end{equation*}}
\def\bes{\begin{equation*}}
\def\ees{\end{equation*}}

\def\d{\, \mathrm{d}}

\usepackage{url}

\title[Gibbs states and Brownian models for coexisting haze and cloud droplets]{Gibbs states and Brownian models for coexisting haze and cloud droplets}

\author[Manuel Santos Guti\'errez]{Manuel Santos Guti\'errez}
\address[MSG]{Department of Earth and Planetary Sciences, Weizmann Institute of Science, Rehovot 76100, Israel}

\author[Micka\"el D. Chekroun]{Micka\"el D. Chekroun}
\address[MDC]{Department of Earth and Planetary Sciences, Weizmann Institute of Science, Rehovot 76100, Israel\\
Department of Atmospheric \& Oceanic Sciences, University of California, Los Angeles, CA 90095-1565, USA} 
\email{michael-david.chekroun@weizmann.ac.il}

\author[Ilan Koren]{Ilan Koren}
\address[IK]{Department of Earth and Planetary Sciences, Weizmann Institute of Science, Rehovot 76100, Israel}


\begin{document} 
\begin{abstract}
Cloud microphysics studies include how tiny cloud droplets grow, and become rain. This is crucial for understanding cloud properties like size, lifespan, and impact on climate through radiative effects. Small, weak-updraft clouds near the haze-to-cloud transition are especially difficult to measure and understand. They are abundant but hard to capture by satellites. Köhler's theory explains initial droplet growth but struggles with large particle groups. Here, we present a stochastic,  analytical framework building on Köhler's theory to account for (monodisperse) aerosols and cloud droplets interaction through competitive growth in a limited water vapor field. These interactions are modeled by sink terms while fluctuations in supersaturation affecting droplet growth are modeled by nonlinear, white noise terms.   
Our results identify hysteresis mechanisms in the droplet activation and deactivation processes. Our approach allows for multimodal cloud's droplet size distributions supported by lab experiments, offering a new perspective on haze-to-cloud transition and small cloud formation.


\end{abstract}

\maketitle


\section*{Introduction}
\noindent Clouds play a key role in the climate energy balances and in the water cycle \cite{bengtsson2010global}. A large part of the uncertainties in climate predictions is attributed to limitations in our understanding of cloud physics and therefore, how to parameterize such clouds in global climate models \cite{zelinka2020}. Out of all cloud types, small, warm clouds, are highly abundant and pose serious challenges for cloud research.
 Their diminutive size often falls below the resolution limits of Earth-observing satellites, making direct observation difficult. Additionally, their weak optical signatures can hinder detection and measurement  \cite{koren2007twilight, varnai2009modis, eytan2020}. These factors contribute to a substantial knowledge gap regarding the role and impact of small, warm clouds in the global climate system.

The cloud's droplet size distribution (DSD) is among the most important microphysical properties. The DSD properties are coupled to all dynamical, turbulent, optical, and stochastic cloud processes. DSDs affect the cloud's evolution in time, rain processes, cloud depth, size, and lifetime \cite{khain_pinsky_2018}. When supersaturation is high enough, cloud condensation nuclei (CCNs) are activated into cloud droplets and grow by condensation. This is the main mechanism for droplet growth at the early stages of a warm cloud. If the cloud's dynamics is weak (shallow clouds) or if the aerosol concentration is high, condensation could be the main growth  mechanism throughout the whole cloud's lifetime as droplets do not become large and varied enough to support efficient autoconversion of droplets by the stochastic processes of collision and coalescence; see Chapters 7 and 8 in \cite{rogers1989}.

Köhler theory provides a framework for understanding the activation of aerosol particles into cloud droplets.  In low supersaturation conditions, only a small subset of the aerosol population, typically the largest and most hygroscopic particles, are activated. The activation process is governed by a delicate balance between the Kelvin effect, which opposes activation due to the curvature of the droplet surface, and the Raoult effect, which favors activation due to the presence of dissolved solutes. This interplay is central to Köhler theory \cite{kohler_1936}.
As a result of this competition, aerosol particles can exist in two distinct thermodynamic states: haze particles and activated droplets \cite{khain_pinsky_2018}. Haze particles remain typically in stable equilibrium with the surrounding environment, while activated droplets grow by diffusion. However other microphysics regimes in a turbulent flow  can lead to intricate interactions between haze and activated droplets, resulting e.g.~into nonlinear behaviors such as oscillatory pulses  
of activation at low particle concentrations \cite{yang_2024}.

The mechanisms of activation and deactivation are, however, not symmetric. In order to activate a cloud droplet, it is necessary to attain a critical level of supersaturation $\lambda_K$, typically well above $0\%$. 
 When the supersaturation is decreased to $\lambda_K$ during the deactivation process, an activated cloud droplet may remain sufficiently large to continue experiencing diffusion of water vapor onto its surface. This positive feedback mechanism results into a deactivation threshold that is strictly lower than $\lambda_K$. In this case, activation-deactivation exhibits hysteresis, a phenomenon previously documented in the cloud physics literature \cite{Korolev2003,arabas2017}.

From a nonlinear dynamics perspective, such an hysteresis phenomenon results from the presence of two saddle-node bifurcation curves that meet tangentially (Chapter 8.2 in \cite{kuznetsovbook}). Recall that a  saddle-node bifurcation is a type of bifurcation where two equilibrium points ``collide'' and disappear as a system's parameter is varied (here supersaturation). It is at the core of the notion tipping points \cite{Ashwin2012} and occurs in many fields of physics such as combustion theory \cite{bebernes2013mathematical,chekroun2018topological}, energy balance  climate models \cite{ghil1976climate,north1981energy}, or ocean dynamics \cite{hawkins2011bistability,weijer2019stability,chekroun2023optimal,Lucarini_Chekroun2023}. 
In the context of cloud physics, the authors in \cite{arabas2017} have shown that the activation process of a cloud particle following the Köhler condensational growth equation,  undergoes such a saddle-node bifurcation in which the haze equilibrium merges with an unstable equilibrium.  At this bifurcation point, the stability of the haze equilibrium is lost due to this merging.  On analytical and numerical grounds  it was shown in \cite{arabas2017} that hysteresis does occur for air-parcel models accounting for temperature and pressure variations, suggesting the presence of the other saddle-node bifurcation where the activated equilibrium merges with an unstable one. In this work, we provide a general framework for hysteresis to take place from the Köhler condensational growth equation alone by incorporating a sink term accounting e.g.~for the evolution of ambient heat and moisture content, or sedimentation effects.

Our approach also accounts for turbulent effects. Clouds in turbulent environments exhibit a complex interplay between microphysical processes and turbulent dynamics. Turbulence strongly influences the life cycle of clouds, as demonstrated in \cite{shaw_2003}. Turbulent models are essential for understanding droplet size spectra broadening \cite{sardina_2015, grabowski2017}, the mixing of cumulus clouds with their surroundings \cite{abade2018,eytan2020}, and the parameterization of mixed-phase clouds in global circulation models \cite{Field2014,furtado2016}.
In warm clouds, turbulence-induced localized temperature gradients and small-scale fluctuations can lead to supersaturation variations. These variations can activate haze particles even in environments with a mean subsaturation, and similarly, cloud particles may deactivate in saturated conditions if the fluctuations are sufficiently strong \cite{prabhakaran2020,abade2018}. Therefore, microphysical fluctuations are crucial for accurately representing droplet size spectra in turbulent cloud environments. We represent these effects by means of stochastic parameterizations involving Brownian motions \cite{csg11,Lucarini_Chekroun2023} and possibly depending on the particle's size.

 Thus, this research explores the activation and deactivation of cloud droplets in warm clouds, offering new theoretical perspectives. We aim to find analytical formulas for describing the cloud's DSDs when both small cloud droplets and haze particles coexist, assuming condensation as the only growth mechanism. 
To this end,   we add a sink term to the classical formulation of K\"ohler's equation in order to parameterize the interaction of a population of droplets or other size-depleting mechanisms, like water vapor consumption or sedimentation \cite{arabas2017,hirsch2015properties,krueger2020}. The resulting equation  describing condensational growth of droplets composing a monodisperse aerosol population  allows for multiple stable states to co-exist besides the ``haze" state, meaning the system exhibits multistability.   Although the haze state remains stable, this multistability suggests that activated droplets can get stuck around a specific size due to the balancing act between two factors: the Köhler instability and mechanisms that decrease supersaturation or remove particles; see also \cite{Korolev2003,arabas2017,chandrakar2019}.

This is where turbulence effects throw a wrench in the picture. The stable states become ``fuzzy states'' that are encountered statistically  (metastable) at various rates of occurrence. This means that particles can switch between activated and deactivated states, similar to how Brownian particles move around in a bumpy landscape. These turbulent jolts also strongly impact the activation-deactivation process into the form of various possible hysteresis loops; see Section  {\it "Hysteresis effects"} below. As a result, droplets can become activated at lower levels of supersaturation compared to the standard Köhler threshold.

Finally, our framework allows us to derive analytical formulas for the DSDs within a cloud volume, accounting for these stochastic turbulent effects. In particular, our stochastic model predicts that the DSDs in the Pi cloud convection chamber experiments behave like Gibbs states, i.e.~an equilibrium probability distribution that maximizes the system's entropy  subject to constraints on its average energy like in statistical mechanics \cite{jordan1998variational}. It is shown that our theoretical predictions based on this formalism closely match the actual DSDs observed in these experiments; see Section {\it "Pi-Chamber empirical distributions vs Gibbs states"} below.

\section*{Results}
\subsection*{K\"ohler theory}
\noindent The theory of K\"ohler is the thermodynamical backbone to explain the formation of cloud droplets out of suspended aerosols in the atmosphere. It establishes the critical relative humidity threshold beyond which suspended particles activate and grow by diffusion (condensation) \cite{kohler_1936}. The critical threshold is above saturation (hence, the supersaturation term is used), and it depends on the initial size and chemical composition of the particle in question. These particles are, according to K\"ohler, present in two distinct/dichotomical thermodynamic states: The first is known as \emph{haze} and it is observed when the aerosol is humidified, although it remains at equilibrium with its moist environment. The second state is referred to as \emph{activated droplet}, whereby the particle harvests the available moisture by diffusion and, in principle, grows indefinitely. However, in nature, the available water vapor molecules that are in supersaturation are limited, and in many cases, droplets and haze coexist in the cloud, and particles can oscillate between the two states. Particularly, in cases of small, warm clouds \cite{hirsch2017, altaratz2021environmental}.

\begin{figure*}[htbp]
	\centering
	\includegraphics[width=0.95\textwidth,height=0.25\textwidth]{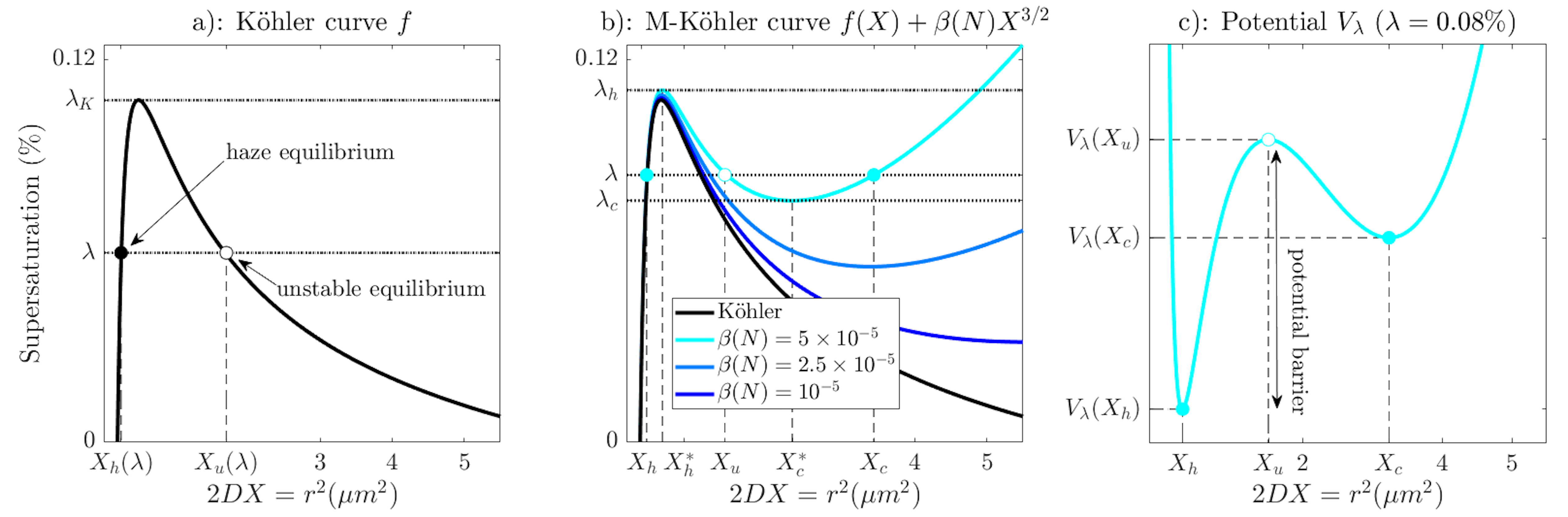} 
	\caption{\label{fig:combo_kohler} {\bf K\"ohler and M-K\"ohler curves, and underlying potential.} {\bf Panel a)}: K\"ohler curve associated with $\mathrm{NaCl}$; see Table \ref{Table_NaCl} for the parameter values.
  A dry radius of $50\mathrm{nm}$ is prescribed. $\lambda_K$ denotes K\"ohler's critical supersaturation  and $\lambda$, an example of supersaturation. {\bf Panel b)}: In black, the K\"ohler curve associated with $\mathrm{NaCl}$ as Panel a). The cyan to blue curves show different  M-K\"ohler curves associated with Eq.~\eqref{eq:sink_arabas} for different values  the parameter $\beta(N)$ as indicated in the legend. For the cyan curve are shown furthermore its intersection  with a given supersaturation $\lambda$ in between $\lambda_c$ and $\lambda_h$. Each filled dot  corresponds to a stable equilibrium while the  empty dot corresponds to the unstable equilibrium. {\bf Panel~c)}: The potential function associated with the cyan M-K\"ohler curve for the supersaturation value as indicated in the title. The solid/empty dots locate the stable/unstable equilibria.}
\end{figure*}

The critical relative humidity threshold needed for activation of a given aerosol with a solubility constant $k$ and a dry radius $r_d$ is analytically obtained using K\"ohler's equation  \cite{kohler_1936, rogers1989}. Here, to simplify  some derivations, $\lambda$  denotes the part of the relative humidity above $100\%$. If $r(t)$ denotes the radius of a given droplet at time $t$, its evolution is approximated by \cite{kohler_1936,arabas2017}:
\begin{equation}\label{eq:kohler}
\begin{cases}
  \dot{r} = \frac{D}{r}(\lambda - \lambda_{eq}(r)) \\
  \lambda_{eq}(r) =  \frac{A}{r} - \frac{B}{r^3}, 
\end{cases}
\end{equation}
where $D$ is the diffusivity constant, $A$ relates to the water surface tension, $B=k r_d^3$, $\lambda$ is the ambient supersaturation and $\lambda_{eq}(r)$ determines the equilibrium supersaturation at the surface of a droplet of radius $r$; commonly known as \emph{K\"ohler curve} \cite{wallace_hobs}. Equation~\eqref{eq:kohler} shows that the particle is in equilibrium when $\lambda = \lambda_{eq}(r)$. To simplify the proposed model, we replace $r$ with the variable $X = r^2/2D$,  which has time-units and could be viewed as relaxation time for diffusion. With this change of variable,   Eq.~\eqref{eq:kohler} translates into the nondimensional form:
\be\label{eq:def_f}
\begin{cases}
 \dot{X} =\lambda-f(X),\\
 f(X)=A(2DX)^{-1/2} - B(2DX)^{-3/2},
 \end{cases}
\ee
with $f(X)$ denoting the equilibrium supersaturation  $\lambda_{eq}(r)$ as  function of (the scaled squared radius) $X$.
A rudimentary analysis of $f(X)$ reveals a single global maximum located at $X_K = 3B/(2DA)$, and with
 critical supersaturation value $\lambda_K = f(X_K) = (4A^3/27B)^{1/2}$, see also Chapter 6.1 in \cite{wallace_hobs}. An example of K\"ohler curve is shown in Fig.~\ref{fig:combo_kohler}a)  whose parameters are given in Table \ref{Table_NaCl}. This curve reaches its maximum at $X_K=6\cdot 10^{-3}\mathrm{s}$, or at $r^2 = 0.48 \mathrm{\mu m}^2$.
\begin{center}
\begin{table}[htbp]
\centering
\caption{\bf K\"ohler curve's parameters for NaCl.}
\renewcommand{\arraystretch}{1.8}
\setlength{\tabcolsep}{4.5pt}
\begin{tabular}{ccccc}
\hline 
 & $k$ &  $r_d$ $(\mu\mathrm{m})$ & $A$ $(\mu\mathrm{m})$ & $D$ ($\mu\mathrm{m}^2s^{-1}$)  \\ 
$\mathrm{NaCl}$ & $1.28$ & $5\times 10^{-2}$ & $10^{-3}$ & $40$  \\
\hline 
\end{tabular}
\label{Table_NaCl}
\end{table}
\end{center}

For a specified dry radius, $r_d$, a particle will activate into a cloud droplet when the ambient supersaturation, $\lambda$, surpasses its critical supersaturation, $\lambda_K$. This activation process occurs through diffusional growth. Conversely, for a given supersaturation less than $\lambda_K$, particles with radii smaller than $(2DX_u)^{1/2}$ will remain in a stable haze state. These equilibrium points are visually represented by the intersection of a horizontal line at the level of $\lambda$ and the Köhler curve, as depicted in Figure \ref{fig:combo_kohler}a. The solid and empty dots within this figure denote the stable and unstable equilibria, respectively.

While K\"ohler theory is established for single particles, clouds consist of large families of particles coexisting in the same humidity environment. When a collection of droplets consumes the available supersaturation, the particles may cease growing by condensation, and the DSDs stagnate and stop being displaced towards larger sizes \cite{Korolev2003,siewert_bec_krstulovic_2017}. Here we modify the K\"ohler equation to the case of many particles in conditions of small warm clouds, when the available supersaturation is a limit factor.  
The proposed model extends the condensational growth equation---governed by the K\"ohler curve---
by coupling all particles to supersaturation consumption and by adding stochastic disturbances, to account for (micro) small-scale turbulent effects.

\subsection*{Multistable K\"ohler curves}
\noindent 
In cases with relatively small supersaturation, as particles grow, they may consume the available supersaturation and therefore reduce their growth rate until reaching stagnation. In the realm of single-particle models, these effects can be modeled by the introduction of a \emph{sink} function into Eq.~\eqref{eq:def_f}:
\be\label{eq:model 1}
\dot{X}= \lambda - \hspace{0.1cm} \underbrace{(f\left(X\right) - g(X))}_{\mbox{M-K\"ohler curve}},
\ee 
in which $g$ denotes the sink term. The function $f-g$ is called a \emph{M-K\"ohler curve} hereafter. The term $g$ should remain negative and should decrease as $X$ increases, as its main role is for accounting for saturation mechanisms in the particle growth. More precisely, this  sink term in growth rate aims to (i) model the supersaturation  budget, and (ii) parameterize other stagnating mechanisms into the condensational growth rate, for instance, due to sedimentation \cite{krueger2020} or the presence of a thermal inversion layer that reduces supersaturation and stops droplet growth \cite{hirsch2014,altaratz2021environmental}.

The sink term can also relate to first-order fluctuations in supersaturation due to condensation. Indeed, when a sizable amount of particles coexist in the same cloud parcel, the constant supersaturation assumption 
 has to be amended taking account changes in $\lambda$ in the course of time, assuming $\lambda$ to be time-dependent in the first equation of Eq.~\eqref{eq:def_f}. At a first order, 
we are concerned with describing the evolution of $\lambda(t)=\lambda+\lambda'(t)$ with $\lambda'(0)=0$, i.e.~we are seeking for a model of the  rate of change of the fluctuations $\lambda'$ around a given value $\lambda$.  

Following \cite{arabas2017}, the rate of change  $\lambda'$ can be approximated by the ratio  $-\dot{\rho_v}/\rho_{vs}$, in which  $\rho_v$ denotes the ambient vapor density away from the droplet while, $\rho_{vs}$, denotes the vapor density at saturation. Then,  by expressing $\rho_v$ as $\rho_v=4\pi N \rho_w r^3/3$, accounting for the density of 
liquid water $\rho_w$ and the droplet number concentration $N$ of identical (monodisperse) particles  in the cloud parcel, we arrive at:
 \begin{equation}\label{eq:rh_arabas}
    \frac{\d}{\d t} \lambda'=-\beta(N)\frac{\d}{\d t}{\left(X^\frac{3}{2}\right)}, 
\end{equation}
when written in the variable $X=r^2/(2D)$, where the constant $\beta(N)$ is given by 
\be\label{Eq_betaN}
\beta(N) = \frac{2^{7/2}D^{3/2}\pi\rho_w}{3\rho_{vs}}N;
\ee
see Eqns.~(9) and (10) in \cite{arabas2017}. 

By integrating each side of Eq.~\eqref{eq:rh_arabas}, 
we obtain $\lambda'(t)= - \beta(N)X^{3/2}$. When this expression of $\lambda'$ is used in $\lambda(t)=\lambda+\lambda'(t)$ we arrive at $\lambda(t)=\lambda-\beta(N) X^{3/2}$ which leads, from  the first equation in Eq.~\eqref{eq:def_f}, to
the following condensational growth model:
\begin{equation}\label{eq:sink_arabas}
    \dot{X} =\lambda(t)-f(X)= \lambda - f(X) - \beta(N)X^{3/2}.
\end{equation}
Thus, the term $-\beta(N)X^{3/2}$ acts as the sink function $g$ in Eq.~\eqref{eq:model 1}: it is negative and decreases as $X$ increases. For large values of $X$, it saturates the particle growth. A few  M-K\"ohler curves are shown in Fig.~\ref{fig:combo_kohler}b), for different $\beta(N)$-values. 

A careful examination of Eq.~\eqref{eq:sink_arabas} reveals that K\"ohler's critical supersaturation $\lambda_K$ and radius are modified due to the sink function. In fact, the function $f(X) + \beta(N)X^{3/2}$ can now admit a local minimum additional to the maximum exhibited by the K\"ohler curve $f$. These are obtained as zeros of 
the following depressed cubic polynomial equation:  
\begin{equation}\label{eq:polynomial}
    -3\beta(N)X^3 + A(2D)^{-1/2}X - 3B(2D)^{-3/2} = 0.
\end{equation}
Sufficient conditions for  Eq.~\eqref{eq:polynomial} to have two positive solutions are easily derivable.
Actually, the existence of haze and activated steady states, $X^{\ast}_h$ and $X^{\ast}_c$   (with $X^{\ast}_c\geq X^{\ast}_h$),   can be inferred for condensational growth models analogue to Eq.~\eqref{eq:sink_arabas} with more general sink functions  $-\beta(N)X^{\alpha}$ ($\alpha>1$); see Proposition~\ref{Prop1} in Section {\it "Multistability"} in {\it Material and Methods}.
 In any event,  as shown in Fig.~\ref{fig:combo_kohler}b) the locations of $X^{\ast}_h$ and $X^{\ast}_c$ depend on the strength of the sink function's coefficient $\beta(N)$.

We define then the following supersaturation values
 $\lambda_h = f(X^{\ast}_{h}) - g(X^{\ast}_{h})$, corresponding to the local maximum of $f-g$, and 
$\lambda_c = f(X^{\ast}_{c}) - g(X^{\ast}_{c})$, corresponding to its local minimum. In particular, $\lambda_h \geq \lambda_c$.

\begin{figure}[t]
		\centering
		\includegraphics[width=0.8\textwidth,height=0.45\textwidth]{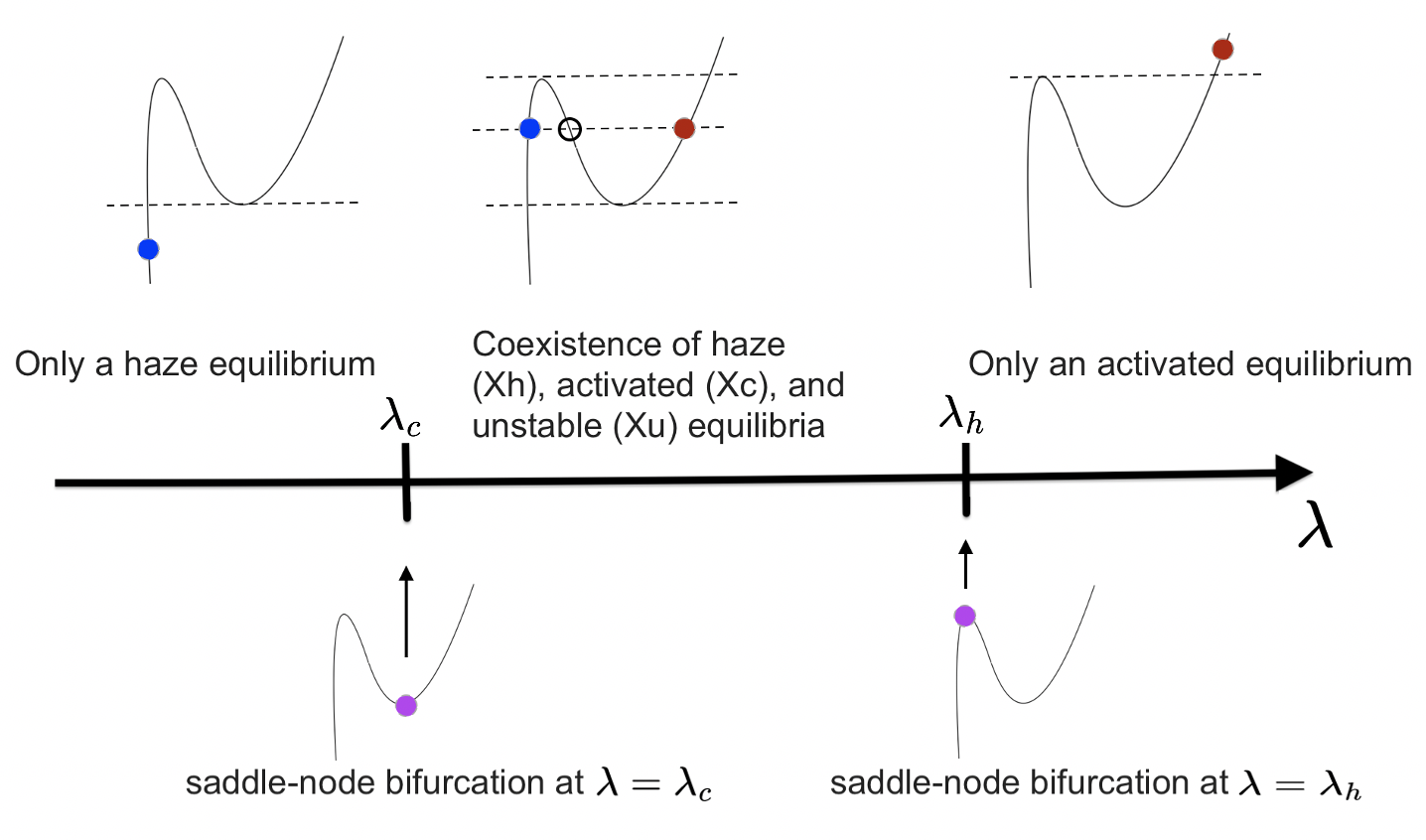}
		\caption{{\bf Going through multiple equilibria as $\lambda$ varies}. Two saddle-node (SN) bifurcations occur, one at $\lambda=\lambda_c$ and one $\lambda=\lambda_h$. For $\lambda<\lambda_c$, there is only a haze stable equilibrium $X_h$ (blue dot in the leftmost upper diagram). As one crosses from below the SN bifurcation point at $\lambda=\lambda_c$, two other equilibria emerge for a total of 3 equilibria coexisting for  $\lambda_c < \lambda <\lambda_h$ (middle upper diagram): an {\it activated} stable equilibrium $X_c$ (red dot), an unstable equilibrium $X_u$ (open circle), and a haze equilibrium (blue dot). At the SN bifurcation occurring at $\lambda=\lambda_h$,  $X_u$ and $X_h$ ``collide'' (purple dot on the  rightmost lower diagram), and only the activated equilibrium survives as $\lambda>\lambda_h$ (red dot in the rightmost upper diagram).} \label{Fig_schematic_SN} 
\vspace{-2ex}		
\end{figure}

Due to the introduction of a new local minimum compared to the original Köhler curve (Fig.~\ref{fig:combo_kohler}a), the number of equilibria varies with changes in supersaturation, $\lambda$, through two saddle-node bifurcations. As depicted in Figure \ref{Fig_schematic_SN}, these bifurcations occur when the haze equilibrium (solid blue dot) approaches and merges with the unstable node (empty circle). Conversely, as $\lambda$ decreases, a similar bifurcation takes place when the activated equilibrium approaches the unstable node. The solid and purple dots in the bottom panels denote the critical values of $\lambda$ at which these bifurcations occur. 
These two saddle-node bifurcations serve as the nonlinear foundation for the hysteresis phenomenon discussed in the following Section on {\it "Hysteresis effects."}
These hysteresis effects are well-known to be revealed in presence of stochastic disturbances combined with slow parameter variations; see e.g.~\cite{berglund_2011}. We discuss below the physical motivation of introducing such disturbances in Eq.~\eqref{eq:sink_arabas} or more generally, in Eq.~\eqref{eq:model 1}.

\subsection*{Brownian models and turbulent effects}

\noindent In a developing cloud, turbulent updrafts and mixing with its neighboring air provoke supersaturation fluctuations that naturally affect droplet growth and, consequently, the collective droplet size spectra. In that respect, we refer to \cite{ditas2012,siebert2017} for observational studies in stratocumulus clouds, and to \cite{paoli2009,KULMALA19971395} regarding the influence of temperature and pressure fluctuations in supersaturation distribution. Stochastic models  for the velocity fluctuations and the
supersaturation fields have been considered in previous studies and have shown various degrees of relevance  \cite{sardina_2015,siewert_bec_krstulovic_2017}. For a cloud parcel subject to turbulent updrafts, the authors in \cite{sardina_2015} prove that activated droplet size variance increases, at short times, proportionally to the square-root of time, analogous to Einstein's diffusion formula. Their analytical findings are contrasted with direct numerical simulations of turbulence obeying the incompressible Navier-Stokes equation. In spite of this diffusion result with diverging variance, in \cite{siewert_bec_krstulovic_2017}, the authors find a statistical steady state for droplet size, with DSDs characterized by exponential tails; see their Eq.~(4.3). Noteworthy is that to achieve convergence to a steady state, a substantial proportion of the droplets must reach evaporation, possibly if, on average, they experience negative supersaturations.

Supersaturation being the driving parameter, we assume that it evolves around a given value $\lambda$ with fluctuations encoded by a stochastic term possibly depending on the particle size. 
The rationale behind this modeling assumption is that larger droplets are expected to have a greater impact on the supersaturation budget than the smaller ones, in a limited water vapor field.
This is what we here referred to as inhomogeneous fluctuations, the inhomogeneous character being dependent on the droplets' size. The type of inhomogeneity referred to here should not be confused thus with that caused by inhomogeneous mixing in clouds, occurring under large entrainment events \cite{Korolev2016}.

These considerations invite us to parameterize the fluctuations of $\lambda$ by means of a size-dependent stochastic term,    $\sigma(X)\dot{W}_t$, leading to the following stochastic model:
\begin{equation}\label{eq:model_s}
    \dot{X}= \underbrace{\lambda - f(X) + g(X)}_{\textstyle{-V_\lambda'(X)}} + \sigma(X)\dot{W}_t,
\end{equation}
where $W_t$ is a Brownian motion (and $\dot{W}_t$ is Gaussian and white). An example  of the $\sigma(X)$ function is given in Section {\it "Pi-Chamber empirical distributions vs Gibbs states"} below.

The function $V_\lambda$ is the potential function that collects the nonlinear, deterministic effects in the model. 
Thus, the solutions $X(t)$ to Eq.~\eqref{eq:model_s}  can be regarded as Brownian particles embedded in the potential $V_\lambda$ defined in Eq.~\eqref{eq:model_s}. The equilibria of the deterministic counterpart of Eq.~\eqref{eq:model_s} (i.e.~Eq.~\eqref{eq:model 1}) correspond exactly to the local critical points of the potential function---local minima in Fig.~\ref{fig:combo_kohler}c)---and their stability is determined by the local curvature of $V_\lambda$ at those points, see e.g.~\cite{berglund_2011,pavliotisbook2014}. Note that Eq.~\eqref{eq:model_s} can always be recast into a form in which the stochastic disturbance term becomes $X$-independent, at the expense of changing the potential function via the Lamperti transformation;  see Chapter~3.6 in \cite{pavliotisbook2014}. This trick is recalled in Section  {\it "Lamperti transformation"} in {\it Material and Methods} as it is also part of our model's analysis below.

 For a family of monodisperse droplets in the same cloud volume, each one experiences different supersaturation fluctuations, albeit with the same mean and variance, as encapsulated by the term $\sigma(X)\dot{W}_t$ in Eq.~\eqref{eq:model_s}. When each supersaturation's stochastic realization is averaged over a large number of particles, the Fokker-Planck Equation (FPE) associated with Eq.~\eqref{eq:model_s} provides the DSD at a  given time $t$; see e.g. \cite{pavliotisbook2014,risken}. The stationary solution to the FPE is then given analytically and known as the \emph{Gibbs state} (see \cite{jordan1998variational} and Eq.~(4.35) in \cite{pavliotisbook2014}):
\begin{equation}\label{eq:inv_mes}
\rho_\lambda (X) = Z_{\lambda}^{-1}\frac{\exp\bigg(2 \int^X \frac{\lambda - f(x)+g(x)}{\sigma^2(x)}\d x\bigg)}{\sigma^2(X)},
\end{equation}
where $Z_{\lambda}$ is a normalizing constant. This distribution is a generalization of Weibull distribution found in  Eq.~(6) in \cite{mcgraw2006} to explain diffusive broadening of the DSDs in turbulent clouds. 
Within our framework, it corresponds to Gibbs states without sink function and state-dependency in the noise, i.e.
density distribution $\rho_W$ (for the radius $r$) obtained by setting $f = g \equiv 0$ and $\sigma$ constant.  In this particular case,  the resulting DSD is integrable only if there is mean subsaturation, $\lambda < 0$ and is given by:
\begin{equation}
	\rho_W(r) = Z_\lambda^{-1} r e^{\gamma r^2},
\end{equation}
where $\gamma$ is a parameter proportional to $\lambda$, droplet number concentration and liquid water fraction; see Eq.~(8) in \cite{mcgraw2006}. In comparison, $\rho_\lambda$ in Eq.~\eqref{eq:inv_mes} can be integrable if $\lambda$ is positive and subject to size-dependent fluctuations, provided that the sink function $g$ has suitable decay properties such as detailed in Proposition~\ref{prop:confining}; see Section {\it "The confining potential"} in {\it Material and Methods}. We refer to \cite{chandrakar2019} for other theoretical approaches to infer integrable DSDs for fluctuating supersaturation with size-dependent removal terms.

Our framework, for a broad class of functions $f$, $g$ and $\sigma$, allows for Gibbs state $\rho_\lambda$ that exhibits multimodality intimately related to  shape of the potential $V_\lambda$. More precisely, the Gibbs state's modes, the local maxima of $\rho_\lambda$, are found by solving the following equation:
\begin{equation}\label{eq:mult_eq0_main}
    -V'_\lambda(X) = \frac{1}{2}\sigma ' (X)\sigma(X).
\end{equation}
See Section {\it "Noise-induced metastability"} in {\it Material and Methods} for the derivation of this equation. 
This fundamental equation teaches us how the Gibbs state modes result from the interaction between the noise term and the potential function. The Gibbs state's modes correspond to metastable states for which the random fluctuation term, $\sigma(X)\dot{W}_t$, in Eq.~\eqref{eq:model_s} acts as a source for the particles to experience transitions from a statistically typical droplet size to another one.

When the noise is state-independent, i.e.~when $\sigma(X) = \sqrt{2\varepsilon} $, for some positive constant $\varepsilon>0$ (controlling the noise variance), then  $\sigma'(X)=0$ and the metastable states are directly determined by the zeros of $-V_\lambda'$.  In this case, for Eq.~\eqref{eq:sink_arabas}, the Gibbs states are {\it bimodal} when $\lambda_c < \lambda < \lambda_h$,  and unimodal when there is just one underlying stable equilibrium, i.e.~for either  $\lambda < \lambda_c$ or $\lambda>\lambda_h$, {as shown in Fig.~\ref{Fig_schematic_SN}, whereby the  solid points denote the presence of equilibria, ranging from a single haze equilibrium to an activated one}.

\begin{figure}[htbp]
		\centering
		\includegraphics[width=0.44\textwidth,height=0.44\textwidth]{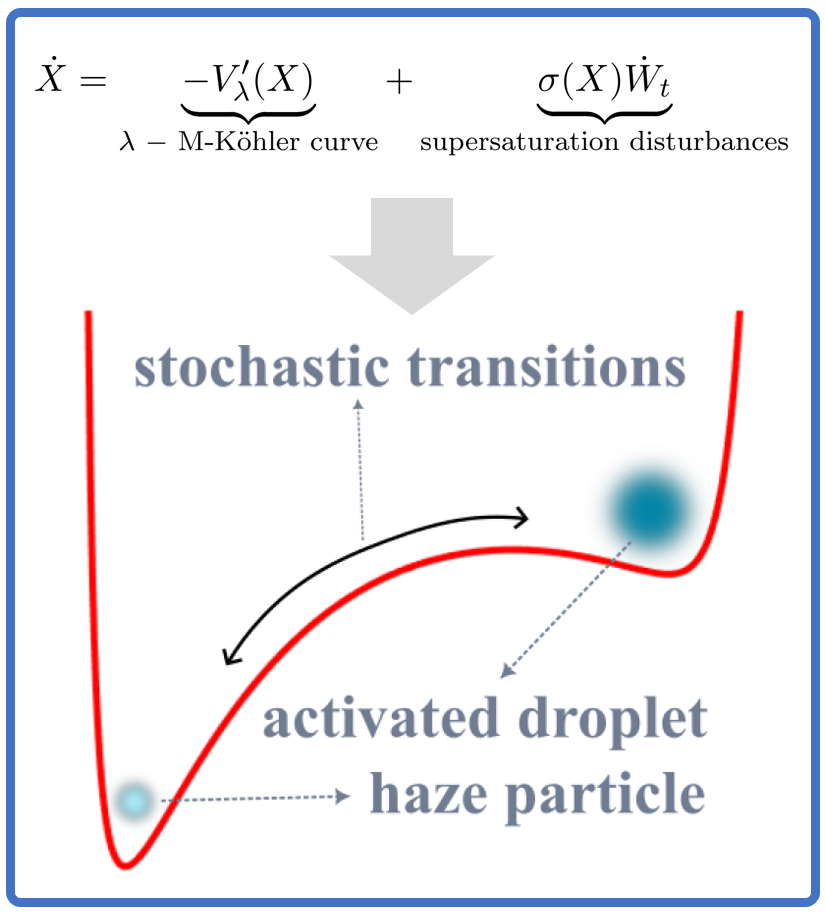}
		\caption{\label{fig:cartoon} {\bf Cartoon of haze-to-droplet stochastic transitions}. Here $\lambda$ is kept fixed and  the situation depicted corresponds to $\lambda$ in $(\lambda_c ,\lambda_h)$ where two stable states coexist  (haze particle and activated droplet); see Fig.~\ref{Fig_schematic_SN}.}
\vspace{-2ex}		
\end{figure}

When $\lambda_c < \lambda < \lambda_h$, there is a notable chance of encountering haze or activated droplets. 
As depicted in Fig.~\ref{fig:cartoon}, the red curve represents a potential energy landscape with two local minima similar to that shown in Fig.~\ref{fig:combo_kohler}c).
 Stochastic fluctuations can then overcome the potential barrier between these two minima, leading to activation even when average supersaturation falls below the Köhler threshold. This phenomenon aligns with real-world scenarios where turbulent supersaturation variations can induce haze activation  \cite{prabhakaran2020}.

Our modeling framework allows for other useful analytic insights. For instance, the  expected residence times at the haze or activated state can be precisely determined. They are indeed known to relate to the potential barrier, $ V_\lambda(X_u) - V_\lambda(X_h)$ (see Fig.~\ref{fig:combo_kohler}c), according to the  Kramers' time formulas \cite{KRAMERS1940284,berglund_2011}. The mean activation time $\tau_{act}$ is then given by Eq.~(2.19) in \cite{matkowski1981}
 \begin{equation}\label{Eq_activate_time}
     \tau_{act}(\lambda) =  \frac{2\pi}{\sqrt{ |V_\lambda^{\prime\prime}(X_u)|V_\lambda^{\prime\prime}(X_h)}}e^{\frac{V_\lambda(X_u) - V_\lambda(X_h)}{\varepsilon}}.
 \end{equation}
Therefore, the closer $\lambda$ is to the critical value $\lambda_h$ from below, the closer $X_u$ is to $X_h$ (see Fig.~\ref{Fig_schematic_SN}) and the shallower the potential barrier becomes, resulting into more common particle activation. Similarly, deactivation of activated droplets  can take place due to  fluctuations in  supersaturation  with a  mean deactivation time obtained by \eqref{Eq_activate_time} in which $X_h$ is replaced by $X_c$.

\subsection*{Hysteresis effects}

\noindent

\begin{figure}[t]
		\centering
            \includegraphics[width=0.85\textwidth,height=0.45\textwidth]{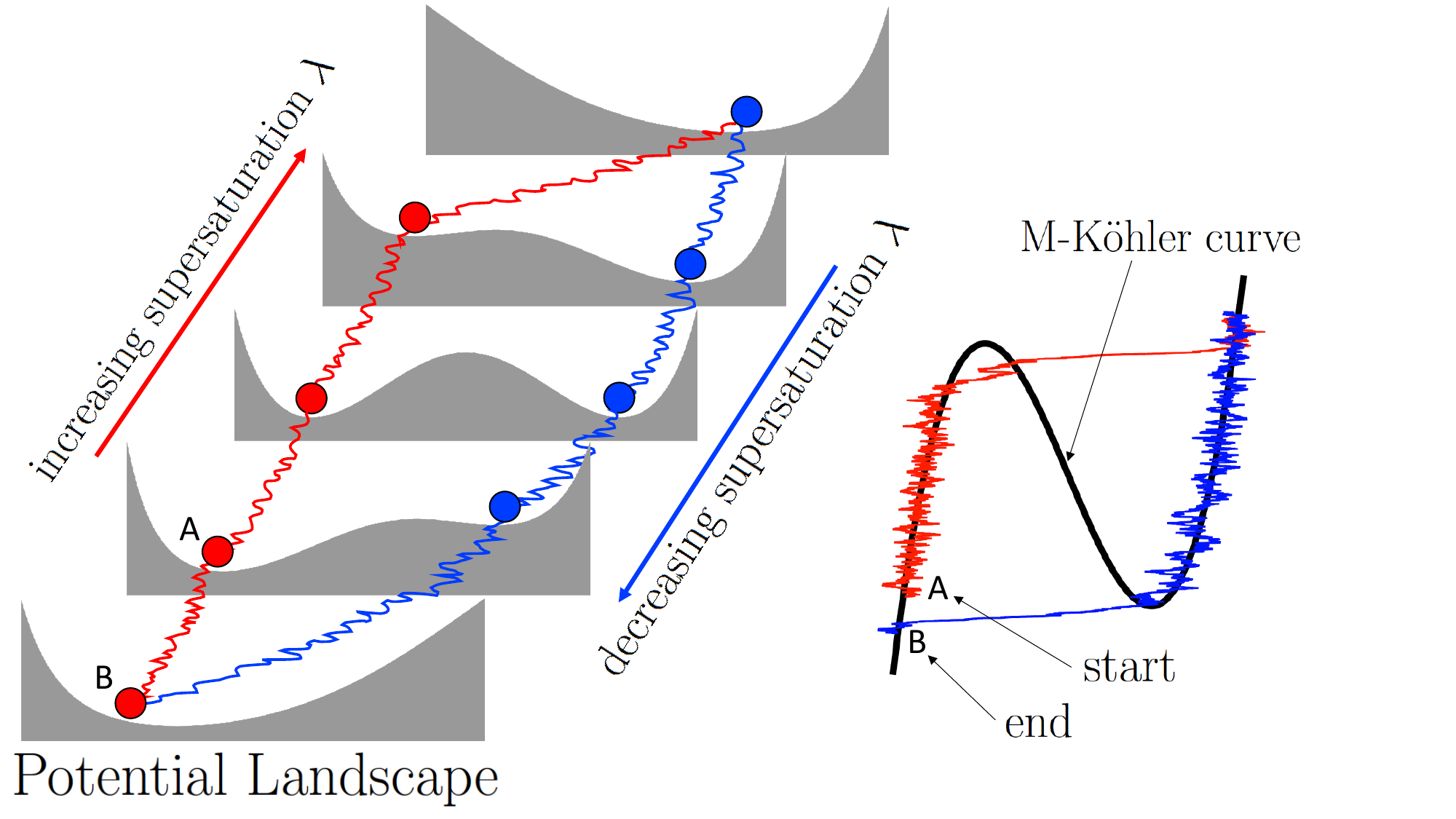}
		\caption{\label{fig:schematic_hysteresis} {\bf Rationale of an hysteresis path}. As $\lambda$ varies, the potential landscape changes and the stochastic disturbances can help trigger the critical transition as one approaches the tipping points, here the saddle-node bifurcation points.}
\end{figure}

\begin{figure*}[t]
    \centering
    \includegraphics[width=0.95\textwidth,height=0.5\textwidth]{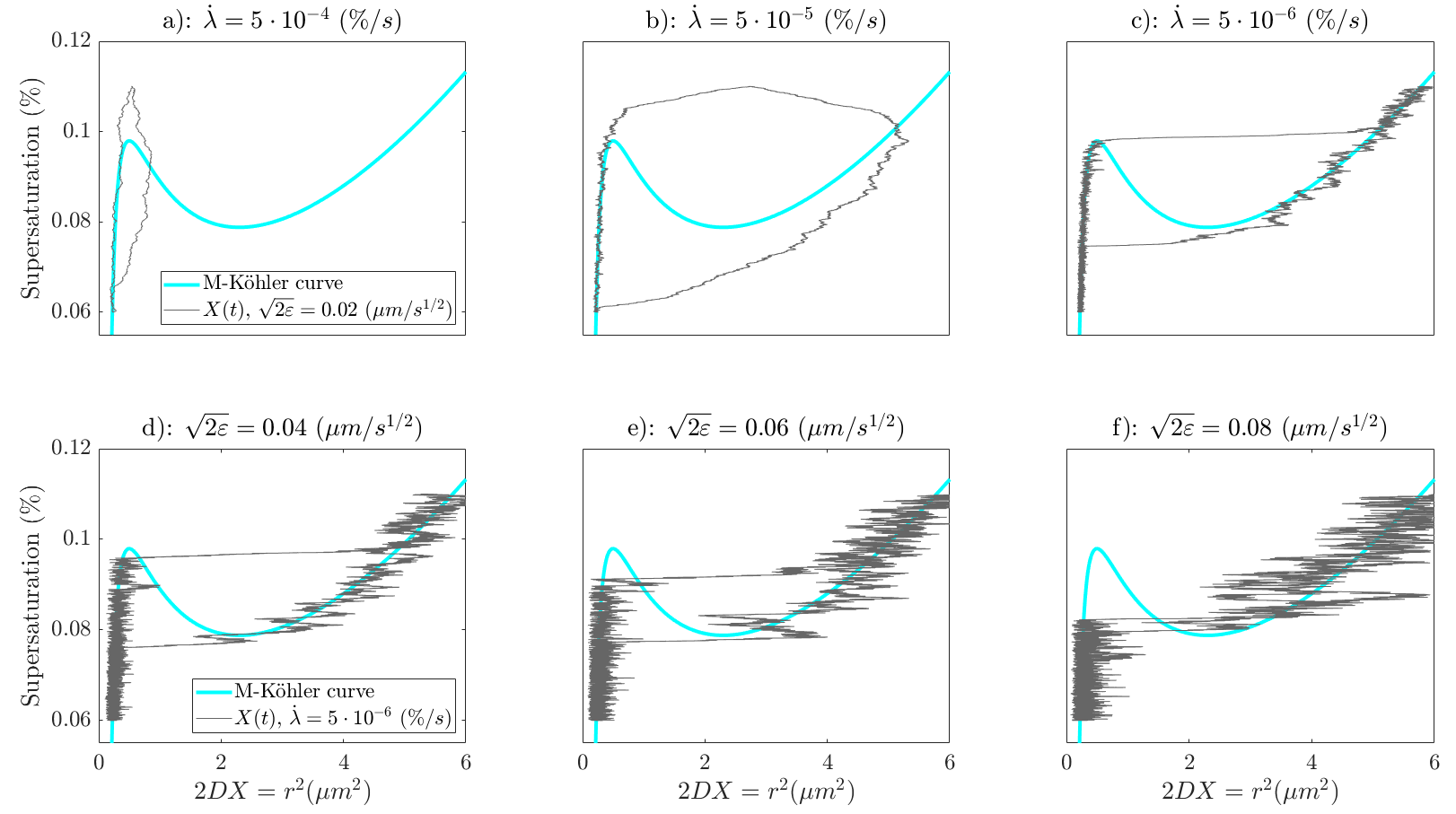}
    \caption{\label{fig:hysteresis_loop} {\bf A landscape of hysteresis loops.} {\bf Panels a), b) and c):} The thin gray line indicates the hysteresis loops for different rates of change of supersaturation $\dot{\lambda}$ as indicated in the title. The noise variance is constant for these three panels as indicated in the legend of Panel a). {\bf Panels d), e) and f):} The thin gray lines indicate the hysteresis loops for different noise variances as indicated in the titles. The rate of change of supersaturation $\dot{\lambda}$ is constant for these three panels as indicated in the legend of Panel d). For all panels, the cyan line is the M-K\"ohler curve associated with Eq.~\eqref{eq:sink_arabas} where $\beta(N) = 3.6\times 10^{-2} \mathrm{s}^{-3/2}$.}
\end{figure*}

\noindent The hysteresis behavior pointed out at the end of Section {\it "Multistable K\"ohler curves"} is now analyzed when the supersaturation parameter $\lambda$ is allowed to drift through the critical values $\lambda_c$ and $\lambda_h$, slower compared to the stochastic fluctuations. Such relatively slower variations in $\lambda$ are associated with buoyancy-driven supersaturation changes. Supersaturation $\lambda$ increases as the buoyant parcel rises   
until the activation threshold $\lambda_h$ is attained and some of the contained haze is activated. When the parcel ceases to be buoyant or reaches a temperature inversion, supersaturation decreases to $\lambda_c$, below which the whole family of droplets regenerates into haze.

During this buoyancy-cycle, the population of particles faces a continuous change in the potential $V_\lambda$, as illustrated in Fig.~\ref{fig:schematic_hysteresis}, for increasing and decreasing supersaturation levels. It is there shown the asymmetry of activation. Indeed, when a haze particle undergoes activation--- starting at point A and following the red, wobbly line---, its deactivation will occur at much lower  supersaturation levels--- as seen when following the blue wobbly line that ends in point B---. Similar cycles have also been reported in the absence of aerosol curvature and chemistry effects, and uniquely due to noisy supersaturation fluctuations around a subsaturated mean; see e.g.~Fig.~4 in \cite{siewert_bec_krstulovic_2017}.

Within the present framework, it remains to investigate the effects of (i) different rates of changes in supersaturation and (ii) variations in the magnitude of the fast supersaturation fluctuations as encapsulated by the noise term. The hysteresis phenomenon under the presence of noise has been investigated in previous work \cite{berglund_2002}, where it is found that a large (additive) noise variance $\sigma$, on average, reduces the area of hysteresis loops. In our context, this implies that droplet activation can occur at much lower supersaturation values compared to $\lambda_h$. To numerically examine these ideas, we solve Eq.~\eqref{eq:model_s} associated with the sink term $g(X) = -\beta(N)X^{3/2}$ used in Eq.~\eqref{eq:sink_arabas}, for  specific noise variances and mean supersaturation $\lambda$ ranging from $0.06\%$ to $0.11\%$ with different rate of change $\dot{\lambda}$. We refer to Section {\it "Hysteresis path algorithm"} in {\it Material and Methods} for numerical details.

Our model supports different hysteresis cycles as shown in Figure \ref{fig:hysteresis_loop}. The nature of these hysteresis loops is controlled by the rate of change of supersaturation $\dot{\lambda}$ and the intensity of the turbulent effects. Figure \ref{fig:hysteresis_loop}a) to  Figure \ref{fig:hysteresis_loop}c) correspond to an increasingly faster response of droplet growth to supersaturation changes. 
Figure \ref{fig:hysteresis_loop}d) to  Figure \ref{fig:hysteresis_loop}f) correspond to an increasing susceptibility to activation as the noise variance is increased, i.e.~the turbulent effects are more pronounced while the rate of change $\dot{\lambda}$ is kept fixed.

As expected, when the rate of change in supersaturation is not slow enough, it can prevent droplet activation--- see Fig.~\ref{fig:hysteresis_loop}a)---, whereas sufficiently slow, adiabatic changes enforce the droplet size to follow the M-K\"ohler curve such as shown  in Fig.~\ref{fig:hysteresis_loop}c). As a haze particle experiences an increase in supersaturation--- resulting from an adiabatic cooling of a cloud parcel--- this will grow by condensation following the K\"ohler curve, almost regardless of how intense the cooling rate is, here indicated by $\dot{\lambda}$. This is an indication of the stability of the thermodynamic equilibrium  of haze particles as opposed to cloud droplets, which are more sensitive the cooling rate $\dot{\lambda}$. Indeed, by comparing Figs.~\ref{fig:hysteresis_loop}b) and c), we observe that only when the cooling rate is slowest do we observe a droplet growth attached to the K\"ohler curve.

Panels d), e) and f) of Figure~\ref{fig:hysteresis_loop} provide for Eq.~\eqref{eq:sink_arabas},  illustrations of hysteresis cycles that are consistent with those analyzed for  large-noise regimes in theoretical studies  \cite{berglund_2002}.
 Compared to Panels a) and b) of Figure~\ref{fig:hysteresis_loop}, the rate of change of supersaturation $\dot{\lambda}$ is smaller, resulting into sharper transitions  to activation similar than those shown in Figure~\ref{fig:hysteresis_loop}c). This combined with a  larger turbulent noise intensity than in the cases of Figure~\ref{fig:hysteresis_loop}a)-b)-c), may provoke transitions to occur before reaching the haze-to-droplet saddle-node bifurcation, i.e.~particles that activate at much lower supersaturations compared to that predicted by K\"ohler. Physically it corresponds to haze particles that grow  in a slowly cooling air volume embedded in a highly turbulent environment undergoing heaving turbulent updrafts. In some dramatic limit, the hysteresis loop may collapse such as shown in Figure~\ref{fig:hysteresis_loop}f). This extreme situation is to put in contrast with the other extreme situation shown in Figure~\ref{fig:hysteresis_loop}a) where activation is "unrealized" due to faster rate of change of supersaturation and weakly turbulent updraft.

\subsection*{Pi-Chamber empirical distributions vs Gibbs states}

\noindent Small, warm convective clouds are abundant in the atmosphere but hard to measure \cite{koren2008small}. Such clouds are often characterized by weak updrafts and small supersaturation, allowing for a substantial part of the aerosol to coexist as haze together with the activated part \cite{altaratz2021environmental, hirsch2017}. Cloud chambers are often used to study clouds in such thermodynamic conditions experimentally \cite{prabhakaran2020, chandrakar2016}.

In such work, the so-called Pi-chamber is seeded with monodisperse families of particles, and turbulent fluctuations are induced by changes in temperature, pressure, and aerosol injection rates. Disregarding transients, the system reaches a steady state in number concentration and DSDs are calculated spanning haze and active particles. Depending on the supersaturation mean and variance, DSDs display different characteristics involving their multimodality and spread;  see Fig.~2 in \cite{prabhakaran2020}. We show here that such experimental distributions are well approximated by the Gibbs states of our Brownian modelling framework once the sink and noise terms are appropriately designed. Our goal is to provide a qualitative and quantitative correspondence between the present theory and the experiments from \cite{prabhakaran2020}.

We first address the sink term design. Note that for a monodisperse family of particles embedded in a cloud parcel, the supersaturation budget must be included in the collective droplet growth \cite{squires_1956,Korolev2003}. Because the Pi-chamber is in steady-state, the sink term affecting supersaturation  is determined by water vapor condensation onto the particles,  which boils down to assuming that supersaturation is consumed proportionally to the radius of the droplets.  Thus a sink term of type $g(X)=-\beta X^{1/2}$ is chosen in Eq.~\eqref{eq:model_s}, where the absorption parameter $\beta$ depends on the number of active droplets, and the effective diffusivity of water vapor in air \cite{khain_pinsky_2018}.   
It must be noted, however, that although in steady-state, the Pi-chamber is not a closed system. There is an intrinsic decorrelation timescale of supersaturation; see \cite{chandrakar2016}.  This timescale is usually determined by the coefficients in front of the deterministic terms of a  (linear) stochastic Langevin equation governing the  supersaturation evolution; see e.g.~Eq.~(2) in  \cite{chandrakar2016}. 

In our formalism,  while not solving an explicit governing equation for supersaturation we can still determine a decorrelation timescale for supersaturation albeit in an indirect fashion. One first solves  Eq.~\eqref{eq:model_s}, records $X(t)$ and then form supersation time series according to:
\be\label{Eq_approx_lambda}
 \lambda(t) =\lambda - \beta X(t)^{1/2} + \sigma(X(t))\dot{W}_t.
\ee 
Now the expectation of the stochastic term 
 $\sigma(X)\dot{W}_t$ is zero due to the martingale property of Brownian motion and the independence of $\sigma(X(t))$ and $\dot{W}_t$. This is actually a fundamental property of stochastic differential equations driven by Brownian motion; see Section 4.2.6 in \cite{gardiner2009}.
 Therefore the expected value of supersaturation has a decorrelation time given by that of $\lambda-\beta \mathbb{E}X(t)^{1/2}$, that is a function of the decorrelation time of the droplet radius. Thus the droplet size effects on supersaturation fluctuations encoded by the stochastic term   $\sigma(X)\dot{W}_t$ affect the decorrelation time of the expected value of supersaturation in an indirect way, after solving Eq.~\eqref{eq:model_s} to get access to $X(t)$.

 In the case of the Pi-chamber datasets analyzed below,  we model the size-dependent effects on supersaturation fluctuations via the following functional:
\bea\label{eta_term_chamber}
&\sigma(X) = \Big(\sigma_1 + \frac{\sigma_2 - \sigma_1}{2}\big(1 + \psi(X)\big)\Big) ,\\
& \psi(X)= \tanh(2 \kappa D (X - X^{\ast})),
\eea
where we require that $\sigma_2>\sigma_1>0$ so that larger droplets have a bigger impact on supersaturation fluctuations than the smaller ones.
The value of the slope is chosen to be sufficiently large --- i.e., $2\kappa D= 10$ --- so that the hyperbolic  tangent function, $\psi$, approximates a step function, while still exhibiting the attributes of a smooth function, more amenable for analysis and numerical simulations.
The ``ignition'' parameter $X^{\ast}$ is chosen to be equal to $6.2\cdot 10^{-3}\mathrm{s}$ or, in diameter, $d^{\ast}=1.41\mu\mathrm{m}$, which yields a correct location of the haze mode, and it is in the order of the critical K\"ohler diameter which is $d_K = 1.7\mu \mathrm{m}$. Such a critical value is derived from Eq.~\eqref{eq:kohler}, as explained in \emph{K\"ohler theory}, for the curvature and solubility coefficients $A = 1.4\cdot 10^{-3}\mu \mathrm{m}$ and $B = 3.5\cdot 10^{-4}\mu \mathrm{m}^{3}$; see \cite{prabhakaran2020}.  

We, thus, arrive at the following Brownian model to describe cloud droplet size evolution in the Pi-chamber:
 \be\label{Eq_Pi_chambermodel}
\dot{X}= \lambda - f(X) - \beta X^{1/2}+ \sigma(X)\dot{W}_t,
 \ee
with $\sigma$ given by \eqref{eta_term_chamber}, and we recall that $f$ is given by \eqref{eq:def_f}.

By solving Eq.~\eqref{eq:mult_eq0_main} associated with Eq.~\eqref{Eq_Pi_chambermodel}, we can derive a relationship (valid for $X$ sufficiently bigger than $X^{\ast}$) determining the parameter $\beta$ in terms of  the modal size $X_c$ of the activated mode:
\begin{equation}\label{eq:beta_estimate}
\beta = \frac{1}{X_c^{2}}\left( \lambda X_c^{3/2} - \frac{A}{(2D)^{1/2}}X_c + \frac{B}{(2D)^{3/2}} \right).
\end{equation}
Eq.~\eqref{eq:beta_estimate} plays a key role to estimate $\beta$: once $\beta$ is known, the only parameters left for calibration of the model are $\sigma_1$, $\sigma_2$ and the stiffness parameter $\kappa$.

\begin{figure*}[ht]
	\centering
	\includegraphics[width=0.95\textwidth,height=0.5\textwidth]{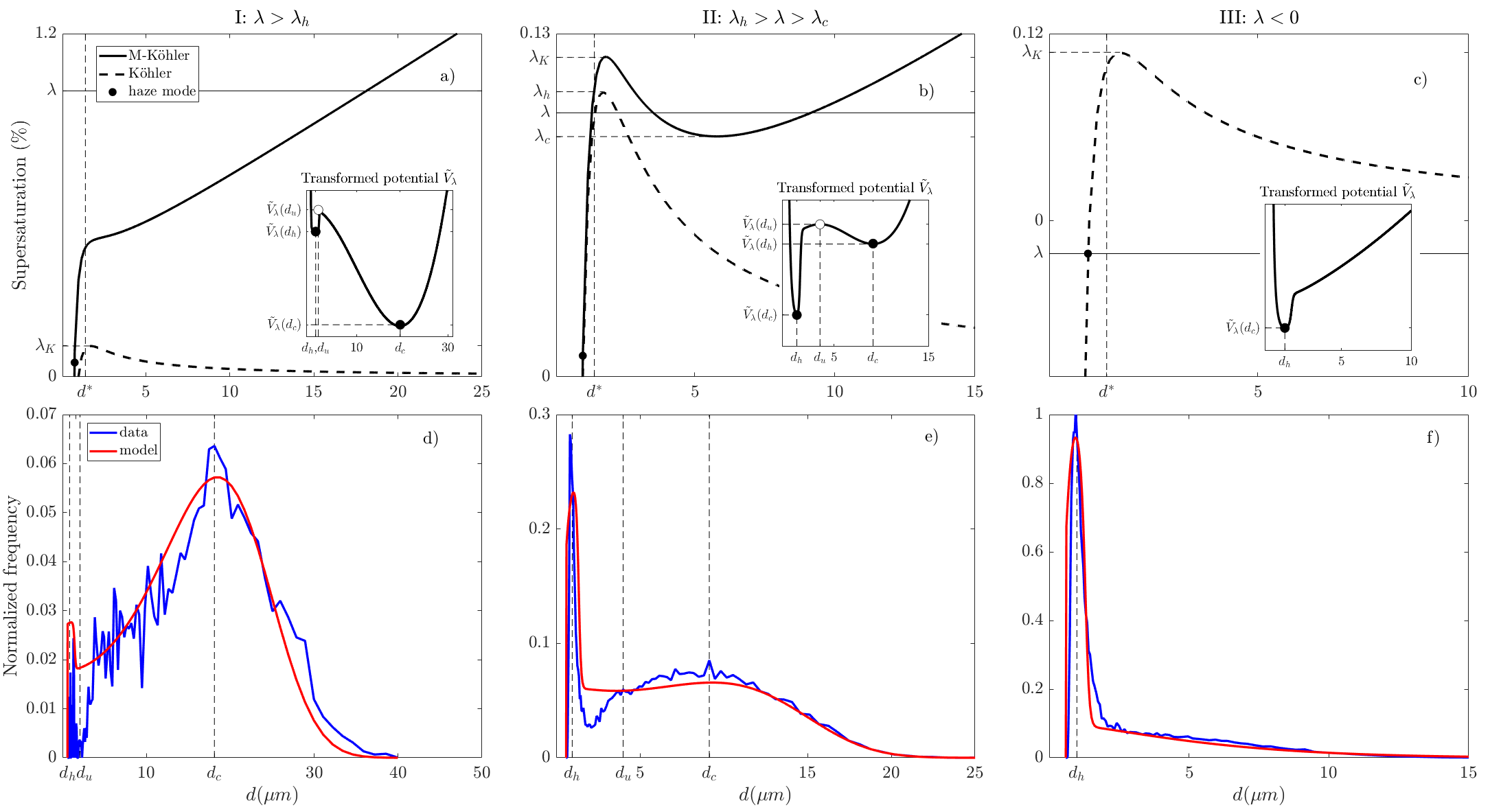}
	\caption{\label{fig:comparison} {\bf Pi-chamber experimental distributions vs Gibbs states, and underlying M-K\"ohler curves.} \textbf{Panels a), b) and c):} K\"ohler and M-K\"ohler curves (dashed and solid  curves, respectively) for each supersaturation value considered in Section {\it "Pi-Chamber Empirical Distribution vs Gibbs States"} (Cases \textbf{I}, \textbf{II}, and \textbf{III}). The parameters  of the K\"ohler curve are $A=1.4\cdot 10^{-3}\mathrm{\mu m}$ and  $B = 3.5\cdot 10^{-4}\mathrm{\mu m^3}$ and the other parameters of model Eq.~\eqref{Eq_Pi_chambermodel} are given in Table \ref{Table_param3}. The black dot denotes the location of the metastable state induced by the state-dependent noise; see Section {\it "Noise-induced metastability"} in {\it Material and Methods}. \textbf{Panels d), e) and f):} Experimental DSDs from \cite{prabhakaran2020} vs Gibbs states from Eq.~\eqref{eq:inv_mes}. Both the data histograms and Gibbs states are normalized according to their integral.  The insets show the transformed potentials (due to Lamperti transformation) explaining the shapes of the Gibbs states; see Section  {\it "The Lamperti transformation"} in {\it Material and Methods}.} 
\end{figure*}

We now confront the ability of our Brownian model \eqref{Eq_Pi_chambermodel} to approximate experimental DSDs by the Gibbs states formula \eqref{eq:inv_mes}. Three experimental setups are considered in that respect, following those of  the Pi-chamber experiments from \cite{prabhakaran2020} corresponding to different supersaturation mean and fluctuation properties reported in such paper. These are referred to as according to \cite{prabhakaran2020}: a) \emph{mean-dominated}, b) \emph{fluctuation influenced} and c) \emph{fluctuation dominated acivation}. 

Within our modeling framework, these regimes can be organized as follows:
\begin{enumerate}
    \item[] \hspace{-3ex}{\bf Case I}: $\lambda > \lambda_K \rightarrow$ mean-dominated activation
    \item[] \hspace{-3ex}{\bf Case II}: $\lambda_c < \lambda < \lambda_K \rightarrow$ fluctuation-influenced activation
    \item[] \hspace{-3ex}{\bf Case III}: $\lambda < 1 \rightarrow$ fluctuation-dominated activation
\end{enumerate}

The parameter $\lambda_K$ is the critical Köhler threshold, which is derived to be $\lambda \approx 1.00108$, for the type of aerosols considered in \cite{prabhakaran2020}. 
The steady-state supersaturation value in each case is also derived from  \cite{prabhakaran2020}. Their theoretical calculations reveal indeed that Case \textbf{I} corresponds to $\lambda = 1\%$. While the value for Case \textbf{II} is not explicitly provided, it can be inferred from our framework that $\lambda$ is close to $\lambda_h$ from below since a bimodal DSD is observed, with high peaks in the haze and activated domain: $\lambda = 0.1\%$  is taken for this case. Finally, Case \textbf{III} is a subsaturated regime ($\lambda < 0$)  and as such we choose $\lambda$ to be given by the  Köhler's value $f(X)$ at the haze peak's location  $X=X_h$, this gives $\lambda = -1\%$.

With these parameter values for $\lambda$, the parameter $\beta$ can be estimated for Cases \textbf{I} and \textbf{II} according to Eq.~\eqref{eq:beta_estimate}, in which $X_c$ is chosen to be the experimental DSD peak's location that  corresponds to activated droplets. For Case \textbf{I}, the modal activated diameter is $d_c = 18.109\mu\mathrm{m}$, giving $X_c = 1.025 \mathrm{s}$ and $\beta = 9.7\cdot 10^{-3}\mathrm{s}^{-1/2}$. In Case \textbf{II}, the modal activated diameter is $d_c=9.141\mu\mathrm{m}$, giving $X_c=0.2611 \mathrm{s}$ and $\beta = 1.4\cdot 10^{-3}\mathrm{s}^{-1/2}$. These modal values are obtained from the data of \cite{prabhakaran2020}.   For Case \textbf{III}, though, since the system is subsaturated and haze particles are more numerous---see Table 1 of \cite{prabhakaran2020}---the term $\beta$ is practically zero. The corresponding original K\"ohler and M-K\"ohler curves are shown in Fig.~\ref{fig:comparison}a) and b), as dashed and solid curves, respectively, and shown as functions of the diameter $d=2\sqrt{2DX}$. For Case \textbf{III}, because there is no sink term, only the original K\"ohler curve is shown in Fig.~\ref{fig:comparison}c).

Once the parameters for the relative disturbances,  $\sigma_1$, $\sigma_2$ and $\kappa$ are chosen as listed in Table~\ref{Table_param3}, the corresponding Gibbs states $\rho_\lambda$ given by Eq.~\eqref{eq:inv_mes} exhibit striking skills in reproducing the empirical DSDs in each of the three  Pi-chamber experiments taken from \cite{prabhakaran2020}. The results are shown in the Figs.~\ref{fig:comparison}d), e) and f). The shapes of the Gibbs states are highly correlated to those of the  transformed potentials shown in the insets of Fig.~\ref{fig:comparison}, obtained by the Lamperti transformation; see Section {\it "The Lamperti transformation"} in {\it Material and Methods}.
We observe in particular that the potential wells' locations match the Gibbs states modes' locations. Recall that the latter transformation allows for identifying the effective potential for the stochastic model  \eqref{Eq_Pi_chambermodel} subject to size-dependent stochastic disturbances such  as given by \eqref{eta_term_chamber}. With this effective potential, Eq.~\eqref{Eq_Pi_chambermodel}  can be interpreted as a Brownian particle evolving within that potential whose fluctuations are just driven by a white noise, independently of the size of the particle.

The theory allows us to make striking predictions. For instance, the experimental DSDs' modes are well predicted by
Eq.~\eqref{eq:mult_eq0_main}, as shown in Fig.~\ref{fig:mult_eq} of Section  {\it "Noise-induced metastability"} in {\it Material and Methods}, especially  in Cases \textbf{I} and \textbf{II}, 
where the classical K\"ohler theory is unable to predict an activated mode, since $\lambda>\lambda_K$. 
Noteworthy is the creation of a metastable haze-state in Case \textbf{I} resulting from the interaction between the noise term $\sigma(X)\dot{W}_t$ and the nonlinear drift terms $-f(X)-\beta X^{1/2}$ as revealed by the transformed potential shown in the inset of  Figs.~\ref{fig:comparison}d); see Section  {\it "Noise-induced metastability"} in {\it Material and Methods}.

\begin{center}
\begin{table*}
\centering
\caption{\label{Table_param3} {\bf Supersaturation values, sink term's and state-dependent noise's parameters}.} 
\renewcommand{\arraystretch}{1.8}
\setlength{\tabcolsep}{7pt}
\begin{tabular}{ccccccc}
\hline \hline
  &$\lambda \times 100$ ($\%$)& $\beta$ $(\mathrm{s}^{-1/2})$ &$\sigma_1$ $(\mathrm{s}^{1/2})$ & $\sigma_2$ $(\mathrm{s}^{1/2})$ & $d^{\ast}$ $(\mu \mathrm{m})$ & $2\kappa D$  \\ \hline
 \textbf{Case I} & $1$ & $ 9.6\cdot 10^{-3}$ & $3.75\cdot 10^{-2}$ & $6.25\cdot 10^{-2}$ & 1.41 & 10 \\
 \textbf{Case II} & $0.1$ & $ 1.4\cdot 10^{-3}$& $7.5\cdot 10^{-3}$  & $1.5\cdot 10^{-2}$ & 1.41 & 10 \\
 \textbf{Case III} & $-1$ & $ 0$&$5\cdot 10^{-3}$ & $1.5\cdot 10^{-2}$ & 1.41 & 10 \\
\hline \hline
\end{tabular}
\end{table*}
\end{center}

\subsection*{Discussion}

\noindent 
K\"ohler theory is at the basis of cloud physics. For a given dry aerosol size and composition, it predicts the particle's state, haze, or activated droplet as a function of the particle size and the supersaturation. The K\"ohler curve (Fig.~\ref{fig:combo_kohler}a), 
has two fixed-point regimes bounded by the curve's mode. The curve left to the mode describes a stable fixed point, whereas any particle that resides on the curve right to the mode is on an unstable fixed point such that a perturbation to the right (or upward) will shift the particle to a continuous growth by diffusion and a perturbation to the left (or downward) will send the particle to the haze state. 
Warm and small clouds that are driven by weak temperature or humidity perturbations \cite{eytan2020, altaratz2021environmental, hirsch2017} are abundant and have an important radiative effect. However, due to their sizes, weak optical signature, and short lifetimes, these clouds are mostly overlooked, and their properties are poorly understood. Such clouds often exist near the transition between haze to cloud \cite{koren2009aerosol} a regime that can be studied in cloud chambers \cite{prabhakaran2020, chandrakar2016}. 

Here, we address these clouds by extending K\"ohler theory to a population of droplets embedded in turbulent environments.
While activation is explained by K\"ohler's theory, the reverse process, i.e.~deactivation, is dynamically different. It has been documented in previous works \cite{arabas2017,Korolev2003} and in the present study that deactivation  is intimately tied to the hysteresis phenomenon through which deactivation occurs at a smaller supersaturation compared to the critical K\"ohler value $\lambda_K$. The addition of a sink term in the condensational growth equation --- see Eq.~\eqref{eq:model 1} --- yields hysteresis directly from an analytical perspective: in order to deactivate a family of monodisperse cloud droplets, the supersaturation needs to be brought strictly below the critical value $\lambda_K$ predicted by K\"ohler. The origin of this behavior is the presence of  another saddle-node bifurcation that results from the bending of the M-K\"ohler curve introduced in Section {\it "Multistable K\"ohler curves}." This bending translates further into the appearance of another stable equilibrium on the activated droplet spectrum as a balance between condensational instability and sink factors, like supersaturation consumption or particle removal. 

The introduction of Brownian noise in the condensational growth equation parameterizes the influence of supersaturation fluctuations and gives analytical formulas for computing DSDs, here identified as Gibbs states; see Eq.~\eqref{eq:inv_mes}. Such probability distributions generalize earlier formulas for DSDs found in the literature such as Weibull distributions \cite{mcgraw2006,chandrakar2019,krueger2020}. 

The resulting Droplet Size Distributions (DSDs) are categorized into three types based on the mean supersaturation parameter $\lambda$. In the first type, fluctuations in supersaturation are insufficient to sustain droplet activation for extended periods, leading to a unimodal DSD centered around the haze state. In the second type, activation is possible, and when the threshold $\lambda_h$ is exceeded, haze particles become unstable. This results in a single DSD mode peaking at activated sizes. Notably, even if Köhler's threshold supersaturation is not reached, there is a finite probability of particle activation. The associated activation timescales are linked to the average time required for droplets to overcome the activation energy barrier (Fig.~\ref{fig:combo_kohler}c), as accurately described by Kramers' formula (Eq.~\eqref{Eq_activate_time}).

When a cloud parcel undergoes adiabatic cooling, supersaturation fluctuates around a gradually increasing mean. Ignoring turbulent fluctuations can lead to an overestimation of the activation threshold, as illustrated in Figure \ref{fig:hysteresis_loop}. This figure depicts various hysteresis loops where activation occurs at even lower supersaturation values than $\lambda_h$, especially when the noise variance is large.

Our stochastic M-Köhler model has demonstrated remarkable agreement with steady-state cloud experiments conducted in the Pi-chamber. Notably, it accurately captures the intricate DSD shapes, including bimodal features, as observed in the experimental study by  \cite{prabhakaran2020} across various saturation regimes. To further explore this consistency, future investigations should focus on slow adiabatic changes in mean supersaturation. Based on the analytical understanding of droplet activation-deactivation hysteresis (Fig.~\ref{fig:hysteresis_loop}), we anticipate that similar hysteresis behavior will be observed in the experimental setup of \cite{prabhakaran2020}.

In natural environments, weak clouds often encounter aerosols of varying sizes and compositions. Nevertheless, the low supersaturation levels may act as a selective filter, favoring the activation of larger haze particles. This can create conditions that are not substantially different from our theoretical framework. Additionally, the presence of Brownian motion can introduce variability into the particle distribution, even with initially monodisperse conditions. Future studies will explore scenarios involving polydisperse particles with diverse chemical properties and varying Köhler curves.

\section*{Acknowledgments}
The insights discussed in this work have substantially benefited from the reviewers’ constructive comments, and we express our deep gratitude to them.
\paragraph*{Funding:}
This work has been partially supported by  the European Research Council (ERC) under the European Union's Horizon 2020 research and innovation program (grant agreement no. 810370). MSG is grateful to the Feinberg Graduate School for their support through the Dean of Faculty Fellowship.
\paragraph*{Author contributions:}
MDC conceived the presented idea. MSG led the analyses and MDC supported. MSG, MDC, and IK discussed the results and wrote the manuscript. All co-authors provided critical feedback and helped shape the research, analysis and manuscript.
\paragraph*{Competing interests:}
The authors declare no competing interests.
\paragraph*{Data and materials availability:}
All experimental data used in this study are based on the article of \cite{prabhakaran2020} and are publicly available at: {\small \url{https://digitalcommons.mtu.edu/data-files/3/}}.
All data needed to evaluate the conclusions in the paper are present in the paper and/or the Supplementary Materials.


\section*{Material and Methods}
{\small
\subsubsection*{Multistability}
To ensure that the model defined by Equation \eqref{eq:model 1} exhibits multiple stable equilibria,
 it suffices for the M-Köhler curve $f-g$ to possess at least one local minimum and one local maximum. This condition guarantees that the roots of the equation $\lambda - (f - g) = 0$ correspond to at least two stable equilibria and one unstable equilibrium within a certain range of $\lambda$ values.

To simplify the presentation of the results stated in Proposition \ref{Prop1} below,
we introduce the following parameters:
\begin{subequations}\label{eq:defs_tilde}
    \begin{align}
        \tilde{A} &= \frac{A}{(2D)^{1/2}},\\
        \tilde{B} &= \frac{B}{(2D)^{3/2}}.
    \end{align}
\end{subequations}
We then have the following sufficiency conditions for the existence of a local maximum and a local minimum:
\begin{proposition}\label{Prop1}
Let $f$ be the function given in Eq.~\eqref{eq:def_f} and let $g:\mathbb{R}^+ \longrightarrow \mathbb{R}^-$ be a function of the form $g(X)=-\beta X^{\alpha}$, where $\beta>0$ and $\alpha>1$. If $\dot{g}(5\tilde{B}/\tilde{A})>\dot{f}(5\tilde{B}/\tilde{A})$, the M-K\"ohler curve in Eq.~\eqref{eq:model 1} has at least a local maximum and a local minimum.
\end{proposition}

\noindent{\it Proof.}  
To demonstrate the existence of two zeros for the function $\dot{f}-\dot{g}$, it is sufficient to prove that it undergoes at least two sign changes. For $X>0$, this function is given by:
\begin{equation}\label{Eq_toto}
	\dot{f}(X)-\dot{g}(X)=-\frac{\tilde{A}}{2}X^{-3/2} + \frac{3\tilde{B}}{2}X^{-5/2} + \alpha\beta X^{\alpha-1}.
\end{equation}
When $\alpha = 3/2$ as in Eq.~\eqref{eq:sink_arabas}, one recovers that finding the zeros of $\dot{f}-\dot{g}$ is equivalent to finding the roots of the polynomial equation~\eqref{eq:polynomial}. 

First of all, $\dot{f}$ has a single local and global minimum. Indeed, this can be observed  by examining $\ddot{f}$:
\begin{equation}
	\ddot{f}(X) = \frac{3}{4}\tilde{A}X^{-5/2}-\frac{15}{4}\tilde{B}X^{-7/2},
\end{equation}
which has a unique zero at $X = 5\tilde{B}/\tilde{A}$, and 
\begin{equation}\label{eq:f_dot}
	\dot{f}\left( \frac{5\tilde{B}}{\tilde{A}}\right)=- \frac{\tilde{A}}{5}\left( \frac{\tilde{A}}{5\tilde{B}}\right)^{\frac{3}{2}}<0.
\end{equation} 
Furthermore, the two limits below hold:
\begin{subequations}
\begin{align}
	&\lim _{X\rightarrow 0^+}\dot{f}(X) = +\infty \label{eq:limit_kohler_1} \\
	&\lim _{X\rightarrow \infty}\dot{f}(X) = 0. \label{eq:limit_kohler_2}
\end{align}
These two limits together with Eq.~\eqref{eq:f_dot} indicate that $X = 5\tilde{B}/\tilde{A}$ is a global minimum.
\end{subequations}

Because of the smoothness of $f$ and $g$ on $\mathbb{R}^+$, the limit in Eq.~\eqref{eq:limit_kohler_1}, and--- by hypothesis--- $\alpha>1$, we have:
\begin{equation}
	\lim _{X\rightarrow 0^+}\dot{f}(X)-\dot{g}(X) = +\infty,
\end{equation}
there exists $X_1>0$ such that $\dot{f}(X_1)-\dot{g}(X_1)>0$. Moreover, by hypothesis, $\dot{f}(5\tilde{B}/\tilde{A})-\dot{g}(5\tilde{B}/\tilde{A})<0$, and that gives the first change of sign of the function $\dot{f}-\dot{g}$. 

Lastly, by assumption we have that $\alpha>1$, so the following limit holds 
\begin{equation}
	\lim_{X\rightarrow \infty}\dot{g}(X) = \lim_{X\rightarrow \infty}  -\alpha\beta X^{\alpha-1} = -\infty.
\end{equation}
Considering this limit and the one in Eq.~\eqref{eq:limit_kohler_2}, there exists $X_2>5\tilde{B}/\tilde{A}$ such that $\dot{f}(X_2)-\dot{g}(X_2)>0$. This gives the second change of sign. 

Due to the continuity and smoothness of the functions $f$ and $g$ on $\mathrm{R}^+$, the function $\dot{f}-\dot{g}$ must have at least two zeros.

\subsubsection*{The confining potential}

In order for a gradient system to possess a Gibbs state, it is necessary that the associated potential is \emph{confining} see e.g.~Chapter~4 in \cite{pavliotisbook2014}. We provide here sufficient mathematical conditions by which a sink function $g$ in Eq.~\eqref{eq:model_s} ensures  the confining property. Generally, it amounts to imposing a sufficiently strong decay rate to the sink term.

The potential function $V_\lambda$ is said to be confining if:
\begin{equation}
    \lim_{X \rightarrow 0^+}V_\lambda(X) = \lim_{X \rightarrow \infty }V_\lambda(X) = \infty,
\end{equation}
and if for any $\xi>0$,
\begin{equation}\label{Int_cond}
    \int_{0}^{\infty} e^{-\xi V_\lambda(X)}\d X<\infty.
\end{equation}

Below, we provide a class of sink functions for which the associated M-K\"ohler curve has a confining potential.
\begin{proposition}\label{prop:confining}
    Let $\lambda$ be in $\mathbb{R}$, $f$ be as in Eq.~\eqref{eq:def_f} and let $g:\mathbb{R}^+ \longrightarrow \mathbb{R}^-$ be defined as $g(X) =  -\beta X^{\alpha} $, for $X,\alpha,\beta>0$. Then, the potential $V_\lambda$ associated with Eq.~\eqref{eq:model_s} is confining.
\end{proposition}
\noindent{\it Proof.} By the definition of $f$, $g$ and the potential $V_\lambda$, we find that:
\begin{subequations}
\begin{align}
    -V&_\lambda(X)=    \int^{X}\left(\lambda -f(X) + g(X)\right)\d X \\ &= \lambda X - 2\tilde{A}X^{1/2}-2\tilde{B}X^{-1/2}-\frac{\beta}{1+\alpha}X^{1+\alpha},
\end{align}
\end{subequations}
where the parameters $\tilde{A}$ and $\tilde{B}$ are defined in Eq.~\eqref{eq:defs_tilde}. Then,
\begin{subequations}
\begin{align}
    \lim _{X \rightarrow 0^+}V_\lambda(X) = \lim_{X \rightarrow 0^+} 2\tilde{B}X^{-1/2} =  \infty, \\ 
    \lim _{X \rightarrow \infty}V_\lambda(X) = \lim_{X \rightarrow \infty} \frac{\beta}{1+\alpha}X^{1+\alpha} = \infty.
\end{align}
\end{subequations}

We are left with showing the integrability condition \eqref{Int_cond}. The integral of the exponential of the potential satisfies, for any $\xi>0$, the following inequalities:
\begin{subequations}
\begin{align}
    &\int _{0}^\infty e^{-\xi V_\lambda(X)}\d X \\ &= \int_{0}^{\infty} e^{\xi\left(\lambda X - 2\tilde{A}X^{1/2}-2\tilde{B}X^{-1/2} -\frac{\beta}{1+\alpha}X^{1+\alpha} \right)} \d X\\&\leq \int_0^{\infty} e^{\xi\left(\lambda X -\frac{\beta}{1+\alpha}X^{1+\alpha} \right)} \d X\\&=\int_0^{\infty} e^{-X\left( \frac{\xi\beta}{1+\alpha}X^{\alpha} -\xi\lambda   \right)}\d X \\ &\leq \int_0^{X_0} e^{-X\left( \frac{\xi\beta}{1+\alpha}X^{\alpha} -\xi\lambda   \right)}\d X + \int_{X_0}^{\infty} e^{-X }\d X< \infty,
\end{align}
\end{subequations}
for any $\xi>0$ and where $X_0 = ((1+\xi\lambda)(1+\alpha)/\xi \beta)^{1/\alpha}$ is such that:
\begin{equation}
     \frac{\xi\beta}{1+\alpha}X_0^{\alpha} -\xi\lambda = 1.
\end{equation}
This proves the confining property.

\subsubsection*{Hysteresis path algorithm}

To produce the hysteresis paths shown in Fig.~\ref{fig:hysteresis_loop}, we solve Eq.~\eqref{eq:sink_arabas} for slowly varying values of the supersaturation parameter $\lambda$ and Brownian noise's parameter $\sigma$ taken to be additive, i.e.~$\sigma(X) = \sqrt{2\varepsilon}$. The slow drift of the  supersaturation parameter is organized as follows. We start by dividing an interval $[\lambda_{0},\lambda_{N+1}] \supset (\lambda_c, \lambda_h)$ uniformly into $N+2$ grid points $\{  \lambda_j\}_{j=0}^{N+1}$ such that $\lambda_{j} = \lambda_0 + j \delta \lambda$, with $\delta \lambda=(\lambda_h - \lambda_c)/(N-1)$ and $\lambda_0=\lambda_c-\delta \lambda/2$.  Note that $\lambda_{N+1}>\lambda_h$ and $\lambda_N<\lambda_h$.

 To form the lower branch of this  hysteresis path, we start with $\lambda=\lambda_0$ and from the haze state $X_h(\lambda_0)$, and solve Eq.~\eqref{eq:sink_arabas} using an Euler-Maruyama scheme with time-step  $\delta t =10^{-2}\mathrm{s}$; see Chapter 9.1 in \cite{Kloeden_platen_book}. 
 
 Then $\lambda$ is updated to $\lambda_0+j\delta \lambda$ and Eq.~\eqref{eq:sink_arabas} is iterated over another timestep by still using Euler-Maruyama. The process is repeated $N+1$ times until $\lambda_{N+1}>\lambda_h$. The reverse trajectory is obtained by repeating the algorithm, although starting with $\lambda=\lambda_{N+1}$ at $X_c(\lambda_{N+1})$. More precisions are given in Algorithm \ref{algostyle} and hereafter. 
 
\begin{algorithm}
	\caption{Hysteresis path computation}\label{algostyle}
	\begin{algorithmic}[1]
		\STATE  $\delta t$ and $N$ is prescribed by the user
		\STATE $\delta\lambda = (\lambda_h - \lambda_c)/(N-1)$
		\STATE $\lambda^+_{vals} = \{\lambda_{0}+j\delta\lambda\}_{j=0}^{N+1}$, with $\lambda_0=\lambda_c-\delta \lambda/2$
		\STATE $\lambda^{-}_{vals} = \{\lambda_{N+1}-j\delta\lambda \}_{j=0}^{N+1}$
		\STATE \text{Form the set} $\lambda_{vals} =\{\lambda^+_{vals},\lambda^-_{vals}\}$
		\STATE Let $X$ be a $2(N+2)$-dimensional vector approximating the hysteresis path 
		\STATE We choose the initial datum to be $X_h\left(\lambda_0\right)$, i.e.~$X(1)=X_h\left(\lambda_0\right)$
		\FOR{$j = 0:2N+3$}
		\STATE $\mu = \lambda_{vals}(j+1)$
	        \STATE $\%$ Euler-Maruyama scheme to integrate  Eq.~(3)
		\STATE $X(j+1) = X (j) - \delta t V_\mu(X(j)) + \sqrt{2\varepsilon  \delta t} \zeta_{j}$, where $\zeta_\ell$ are independent normally distributed random variables with mean zero and variance unity 
		\ENDFOR
	\end{algorithmic}
\end{algorithm}

The key parameter to produce panels a), b) and c) of Fig.~\ref{fig:hysteresis_loop} is $\dot{\lambda}$. Note that according to the previous algorithm, $\dot{\lambda} \approx \delta \lambda / \delta t$. Consequently, the larger $N$ is, the smaller $\delta \lambda$, and vice versa.

\begin{figure*}[htbp]
	\centering
	\includegraphics[width=.9\textwidth, height=.35\textwidth]{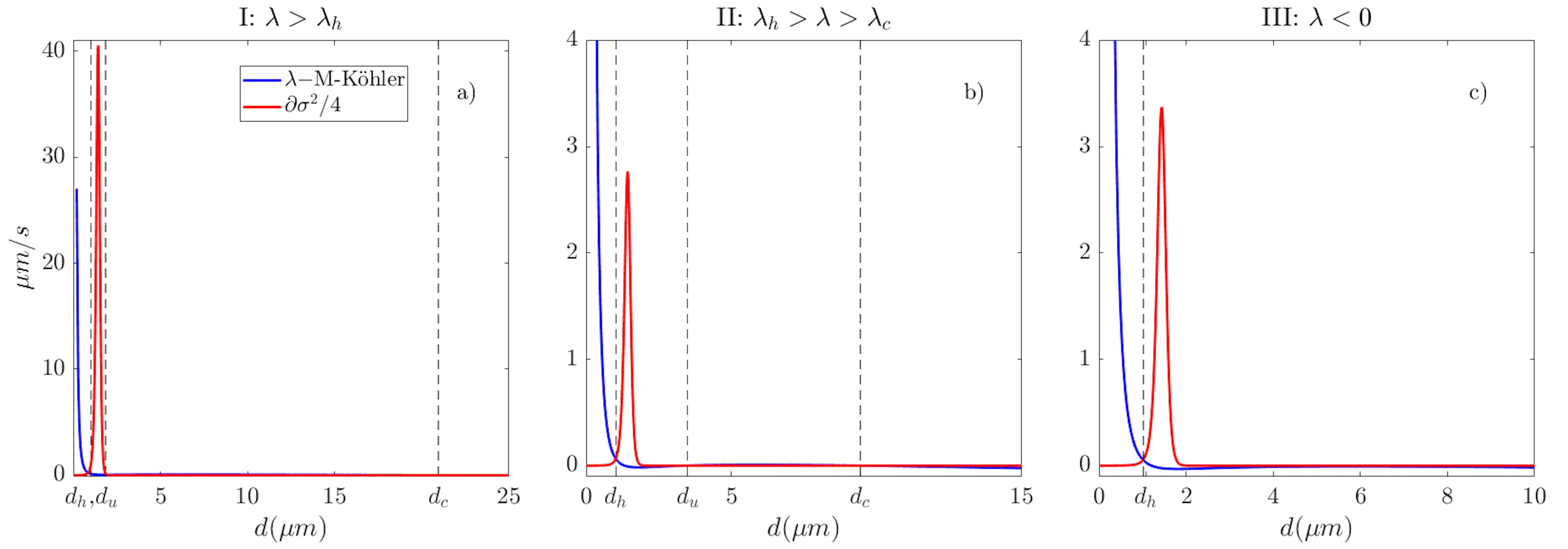}
\caption{ \label{fig:mult_eq} {\bf Noise-induced metastable states}. \textbf{Panels a), b)} and \textbf{c)}, correspond to the three different supersaturation configurations as indicated in the titles and corresponding to those of  Fig.~ \ref{fig:comparison}.  The blue curves show the growth rate function $-V'_\lambda$ in Eq.~\eqref{eq:model_s}, such as associated with the M-K\"ohler curve $f(X)+\beta X^{1/2}$. The red curves show the derivative of the noise covariance after division  by a factor four corresponding to the RHS of Eq.~\eqref{eq:mult_eq0}. The vertical dashed lines indicate the location of the solutions to Eq.~\eqref{eq:mult_eq0} which also correspond to the haze, unstable and activated metastable states, denoted by $d_h,d_u$ and $d_c$, respectively.} 
\end{figure*}

\subsubsection*{The Lamperti transformation}

For scalar stochastic differential model, it is possible to map a stochastic differential equation (SDE) driven by state-dependent noise onto an SDE driven by an additive noise. We recall this basic idea.
Consider the following scalar SDE
\be\label{Eq_Original}
\dot{X}=-V(X)  +\sigma(X) \dot{W}_t,
\ee
with $V$ and $\sigma$ smooth functions, such that $\sigma>0$. 
Consider the change of variables, $h(X)=\int^X \frac{1}{\sigma(X)} \d X$, then by application of It\^o's formula, the original equation \eqref{Eq_Original} is transformed into
\be
\dot{Y}=-\widehat{V}(Y) + \dot{W}_t,
\ee
 with 
 \be\label{Eq_transformed}
\widehat{V}(Y)=\frac{V(h^{-1}(Y))}{\sigma(h^{-1}(Y))}-\frac{1}{2} \sigma'(h^{-1}(Y)).
 \ee
This is called the Lamperti transformation; see Chapter 3.6 in \cite{pavliotisbook2014}.

The Lamperti transformation allows for identifying the {\it effective potential}, $\widehat{V}$, for the stochastic model \eqref{Eq_Original} subject to state-dependent stochastic disturbances given by $\sigma(X) \dot{W}_t$. With this effective potential, Eq.~\eqref{Eq_Original} can be interpreted as a Brownian particle evolving within that potential whose fluctuations are just driven by a white noise, independently of the state $X$.
In particular even if the original potential $V$ does not exhibit a well over a certain range of $X$-values, the transformed potential can exhibit such local minimum. We talk then of noise-induced metastability in case of creation of an extra metastable state resulting from the interaction of the drift term with the diffusion term $\sigma(X) \dot{W}_t$.

\subsubsection*{Noise-induced metastability}

By applying \eqref{Eq_transformed} to our stochastic droplet model \eqref{Eq_Pi_chambermodel} in Pi-chamber, we obtain:
\begin{equation}\label{eq:lamperti_transform}
    \widehat{V}'_\lambda( Y) = -\frac{V'_\lambda(X)}{\sigma(X)}-\frac{1}{2}\sigma ' (X),  \mbox{ with } Y=\int^X \sigma^{-1}(X) \d X,
\end{equation}
in which $V_\lambda'(X)=-\lambda+f(X)+\beta X^{-1/2}.$

Note that the appearance of equilibria non-trivially depends on both the potential $V_\lambda$ and the noise function $\sigma$. Indeed, as a consequence of Eq.~\eqref{eq:lamperti_transform}, the metastable states are explicitly given by solving $\widehat{V}'_\lambda(Y)=0$, namely by solving:
\begin{equation}\label{eq:mult_eq0}
    -V'_\lambda(X) = \frac{1}{2}\sigma '(X)\sigma(X),
\end{equation}
recovering this way Eq.~\eqref{eq:mult_eq0_main}. Thus, if one wishes to match the locations of these metastable states with let us say the modes of an empirical distribution (as in Section {\it "Pi-Chamber empirical distributions vs Gibbs states"}),  one needs to solve Eq.~\eqref{eq:mult_eq0}. As a consequence, this equation imposes constraints on the choice of the parameters involved in $\sigma$ and $V_\lambda$.

 More precisely, we have that \eqref{eq:mult_eq0} rewrites as
\begin{equation}\label{Main_eq_peaks}
\lambda - \tilde{A}X^{-1/2} + \tilde{B}X^{-3/2} -\beta X^{1/2} =  \frac{1}{2}\sigma '(X)\sigma(X),
\end{equation}
where $\tilde{A}$ and $\tilde{B}$ are defined in Eq.~\eqref{eq:defs_tilde}. By observing that $\sigma (X)=\sigma_1 + \frac{\sigma_2-\sigma_1}{2}\big(1+\tanh(\kappa(X-X^{\ast}))\big) \longrightarrow 0$ as $X\rightarrow \infty$,  one can get then, when $\kappa \gg 1$ and  $X$ is sufficiently larger than $X^{\ast}$, good approximations of the solutions $X$ to Eq.~\eqref{Main_eq_peaks} by solving:
\begin{equation}\label{eq:limit_equation}
    \lambda X^{3/2} - \tilde{A}X + \tilde{B} -\beta X^{2} = 0,
\end{equation}
We adopt this strategy for Cases \textbf{I} and \textbf{II} of Pi-Chamber Empirical Distribution vs Gibbs States.

The sink parameter $\beta$ has a direct influence on the location of the metastable states and vice versa. One can invert equations \eqref{eq:limit_equation} to find the value of $\beta$ in terms of the activated mode $X_c$:
\be
    \beta = \frac{1}{X_c^{2}}\left( \lambda X_c^{3/2} - \tilde{A}X_c + \tilde{B} \right),
\ee
leading thus to Eq.~\eqref{eq:beta_estimate}.

The blue and red curves in Fig.~\ref{fig:mult_eq} correspond to the left-hand-side (LHS) and right-hand-side (RHS) of Eq.~\eqref{eq:mult_eq0}, respectively, for  the three case studies of Section {\it "Pi-Chamber empirical distributions vs Gibbs states."}
The intersection of these two curves determines the location of the metastable states indicated by the dashed vertical lines for Cases \textbf{I}, \textbf{II} and \textbf{III} with the parameter configuration shown in Table \ref{Table_param3}. 

}


\newcommand{\etalchar}[1]{$^{#1}$}
\providecommand{\bysame}{\leavevmode\hbox to3em{\hrulefill}\thinspace}
\providecommand{\MR}{\relax\ifhmode\unskip\space\fi MR }
\providecommand{\MRhref}[2]{%
  \href{http://www.ams.org/mathscinet-getitem?mr=#1}{#2}
}
\providecommand{\href}[2]{#2}
\begin{thebibliography}{WCD{\etalchar{+}}19}

\bibitem[AGP18]{abade2018}
Gustavo~C. Abade, Wojciech~W. Grabowski, and Hanna Pawlowska, \emph{Broadening
  of cloud droplet spectra through eddy hopping: Turbulent entraining parcel
  simulations}, Journal of the Atmospheric Sciences \textbf{75} (2018), no.~10,
  3365 -- 3379.

\bibitem[AKA{\etalchar{+}}21]{altaratz2021environmental}
Orit Altaratz, Ilan Koren, Eyal Agassi, Eitan Hirsch, Yoav Levi, and Nir Stav,
  \emph{The environmental conditions behind the formation of small (sublcl)
  clouds}, Geophysical Research Letters \textbf{48} (2021), no.~23,
  e2021GL096242.

\bibitem[AS17]{arabas2017}
S.~Arabas and S.~Shima, \emph{{On the CCN (de)activation~nonlinearities}},
  Nonlinear Processes in Geophysics \textbf{24} (2017), no.~3, 535--542.

\bibitem[AWVC12]{Ashwin2012}
Peter Ashwin, Sebastian Wieczorek, Renato Vitolo, and Peter Cox, \emph{{Tipping
  points in open systems: Bifurcation, noise-induced and rate-dependent
  examples in the climate system}}, Philosophical Transactions of the Royal
  Society A: Mathematical, Physical and Engineering Sciences \textbf{370}
  (2012), no.~1962, 1166--1184.

\bibitem[BE13]{bebernes2013mathematical}
J.~Bebernes and D.~Eberly, \emph{Mathematical problems from combustion theory},
  vol.~83, Springer Science \& Business Media, 2013.

\bibitem[Ben10]{bengtsson2010global}
Lennart Bengtsson, \emph{The global atmospheric water cycle}, Environmental
  Research Letters \textbf{5} (2010), no.~2, 025202.

\bibitem[Ber11]{berglund_2011}
Nils Berglund, \emph{{Kramers' law: Validity, derivations and
  generalisations}}, arXiv preprint 1106.5799 (2011).

\bibitem[BG02]{berglund_2002}
Nils Berglund and Barbara Gentz, \emph{The effect of additive noise on
  dynamical hysteresis}, Nonlinearity \textbf{15} (2002), no.~3, 605.

\bibitem[CCC{\etalchar{+}}16]{chandrakar2016}
Kamal~Kant Chandrakar, Will Cantrell, Kelken Chang, David Ciochetto, Dennis
  Niedermeier, Mikhail Ovchinnikov, Raymond~A. Shaw, and Fan Yang,
  \emph{Aerosol indirect effect from turbulence-induced broadening of
  cloud-droplet size distributions}, Proceedings of the National Academy of
  Sciences \textbf{113} (2016), no.~50, 14243--14248.

\bibitem[Che18]{chekroun2018topological}
M.~D. Chekroun, \emph{Topological instabilities in families of semilinear
  parabolic problems subject to nonlinear perturbations}, Discrete \&
  Continuous Dynamical Systems-B \textbf{23} (2018), no.~9, 3723.

\bibitem[CLM23]{chekroun2023optimal}
M.~D. Chekroun, H.~Liu, and J.~C. McWilliams, \emph{Optimal parameterizing
  manifolds for anticipating tipping points and higher-order critical
  transitions}, {Chaos} \textbf{33} (2023), no.~9, 093126.

\bibitem[CSG11]{csg11}
M~D Chekroun, E~Simonnet, and M~Ghil, \emph{{Stochastic climate dynamics:
  Random attractors and time-dependent invariant measures}}, Physica D
  \textbf{240} (2011), no.~21, 1685--1700.

\bibitem[CSY{\etalchar{+}}20]{chandrakar2019}
Kamal~Kant Chandrakar, Izumi Saito, Fan Yang, Will Cantrell, Toshiyuki Gotoh,
  and Raymond~A. Shaw, \emph{{Droplet size distributions in turbulent clouds:
  Experimental evaluation of theoretical distributions}}, Quarterly Journal of
  the Royal Meteorological Society \textbf{146} (2020), no.~726, 483--504.

\bibitem[DSS{\etalchar{+}}12]{ditas2012}
F.~Ditas, R.~A. Shaw, H.~Siebert, M.~Simmel, B.~Wehner, and A.~Wiedensohler,
  \emph{Aerosols-cloud microphysics-thermodynamics-turbulence: evaluating
  supersaturation in a marine stratocumulus cloud}, Atmospheric Chemistry and
  Physics \textbf{12} (2012), no.~5, 2459--2468.

\bibitem[EKA{\etalchar{+}}20]{eytan2020}
E.~Eytan, I.~Koren, O.~Altaratz, Alex Kostinski, and Ayala Ronen,
  \emph{Longwave radiative effect of the cloud twilight zone}, Nature
  Geoscience \textbf{13} (2020), 669--673.

\bibitem[FFB{\etalchar{+}}16]{furtado2016}
K.~Furtado, P.~R. Field, I.~A. Boutle, C.~J. Morcrette, and J.~M. Wilkinson,
  \emph{A physically based subgrid parameterization for the production and
  maintenance of mixed-phase clouds in a general circulation model}, Journal of
  the Atmospheric Sciences \textbf{73} (2016), no.~1, 279 -- 291.

\bibitem[FHFK14]{Field2014}
P.~R. Field, A.~A. Hill, K.~Furtado, and A.~Korolev, \emph{{Mixed-phase clouds
  in a turbulent environment. Part 2: Analytic treatment}}, Quarterly Journal
  of the Royal Meteorological Society \textbf{140} (2014), no.~680, 870--880.

\bibitem[GA17]{grabowski2017}
Wojciech~W. Grabowski and Gustavo~C. Abade, \emph{Broadening of cloud droplet
  spectra through eddy hopping: Turbulent adiabatic parcel simulations},
  Journal of the Atmospheric Sciences \textbf{74} (2017), no.~5, 1485 -- 1493.

\bibitem[Gar09]{gardiner2009}
Crispin Gardiner, \emph{{Stochastic Methods: A Handbook for the Natural and
  Social Sciences}}, Springer-Verlag Berlin, Heildelberg, 2009.

\bibitem[Ghi76]{ghil1976climate}
M.~Ghil, \emph{Climate stability for a {S}ellers-type model}, Journal of
  Atmospheric Sciences \textbf{33} (1976), no.~1, 3--20.

\bibitem[HKA{\etalchar{+}}17]{hirsch2017}
Eitan Hirsch, Ilan Koren, Orit Altaratz, Zev Levin, and Eyal Agassi,
  \emph{Enhanced humidity pockets originating in the mid boundary layer as a
  mechanism of cloud formation below the lifting condensation level},
  Environmental Research Letters \textbf{12} (2017), no.~2, 024020.

\bibitem[HKAA15]{hirsch2015properties}
Eitan Hirsch, Ilan Koren, Orit Altaratz, and Eyal Agassi, \emph{On the
  properties and radiative effects of small convective clouds during the
  eastern mediterranean summer}, Environmental Research Letters \textbf{10}
  (2015), no.~4, 044006.

\bibitem[HKL{\etalchar{+}}14]{hirsch2014}
E.~Hirsch, I.~Koren, Z.~Levin, O.~Altaratz, and E.~Agassi, \emph{On
  transition-zone water clouds}, Atmospheric Chemistry and Physics \textbf{14}
  (2014), no.~17, 9001--9012.

\bibitem[HSA{\etalchar{+}}11]{hawkins2011bistability}
Ed~Hawkins, Robin~S Smith, Lesley~C Allison, Jonathan~M Gregory, Tim~J
  Woollings, Holger Pohlmann, and B~De~Cuevas, \emph{{Bistability of the
  Atlantic overturning circulation in a global climate model and links to ocean
  freshwater transport}}, {Geophysical Research Letters} \textbf{38} (2011),
  no.~10, L10605.

\bibitem[JKO98]{jordan1998variational}
Richard Jordan, David Kinderlehrer, and Felix Otto, \emph{{The variational
  formulation of the Fokker--Planck equation}}, SIAM Journal on Mathematical
  Analysis \textbf{29} (1998), no.~1, 1--17.

\bibitem[KFJA09]{koren2009aerosol}
Ilan Koren, Graham Feingold, Hongli Jiang, and Orit Altaratz, \emph{Aerosol
  effects on the inter-cloud region of a small cumulus cloud field},
  Geophysical Research Letters \textbf{36} (2009), no.~14, L14805.

\bibitem[KKPF16]{Korolev2016}
A.~Korolev, A.~Khain, M.~Pinsky, and J.~French, \emph{Theoretical study of
  mixing in liquid clouds -- part 1: Classical concepts}, Atmospheric Chemistry
  and Physics \textbf{16} (2016), no.~14, 9235--9254.

\bibitem[KM03]{Korolev2003}
Alexei~V. Korolev and Ilia~P. Mazin, \emph{{Supersaturation of water vapor in
  clouds}}, J.Atmos.Sci. \textbf{60} (2003), no.~24, 2957--2974.

\bibitem[KOF{\etalchar{+}}08]{koren2008small}
Ilan Koren, L~Oreopoulos, G~Feingold, LA~Remer, and O~Altaratz, \emph{How small
  is a small cloud?}, Atmospheric Chemistry and Physics \textbf{8} (2008),
  no.~14, 3855--3864.

\bibitem[K{\"o}h36]{kohler_1936}
Hilding K{\"o}hler, \emph{The nucleus in and the growth of hygroscopic
  droplets}, Trans. Faraday Soc. \textbf{32} (1936), 1152--1161.

\bibitem[KP92]{Kloeden_platen_book}
Peter~E Kloeden and Eckhard Platen, \emph{Numerical solution of stochastic
  differential equations}, vol.~23, Springer Berlin Heidelberg, 1992.

\bibitem[KP18]{khain_pinsky_2018}
Alexander~P. Khain and Mark Pinsky, \emph{Physical processes in clouds and
  cloud modeling}, Cambridge University Press, 2018.

\bibitem[Kra40]{KRAMERS1940284}
H.A. Kramers, \emph{Brownian motion in a field of force and the diffusion model
  of chemical reactions}, Physica \textbf{7} (1940), no.~4, 284--304.

\bibitem[KRK{\etalchar{+}}07]{koren2007twilight}
Ilan Koren, Lorraine~A Remer, Yoram~J Kaufman, Yinon Rudich, and J~Vanderlei
  Martins, \emph{On the twilight zone between clouds and aerosols}, Geophysical
  Research Letters \textbf{34} (2007), no.~8, L08805.

\bibitem[Kru20]{krueger2020}
S.~K. Krueger, \emph{Technical note: Equilibrium droplet size distributions in
  a turbulent cloud chamber with uniform supersaturation}, Atmospheric
  Chemistry and Physics \textbf{20} (2020), no.~13, 7895--7909.

\bibitem[KRZC97]{KULMALA19971395}
Markku Kulmala, {\"U}llar Rannik, Evgeni~L. Zapadinsky, and Charles~F. Clement,
  \emph{The effect of saturation fluctuations on droplet growth}, Journal of
  Aerosol Science \textbf{28} (1997), no.~8, 1395--1409.

\bibitem[Kuz04]{kuznetsovbook}
Yuri~A. Kuznetsov, \emph{Elements of applied bifurcation theory}, Springer,
  2004.

\bibitem[LC23]{Lucarini_Chekroun2023}
Valerio Lucarini and M.D. Chekroun, \emph{{Theoretical tools for understanding
  the climate crisis from Hasselmann's programme and beyond}}, Nature Reviews
  Physics \textbf{5} (2023), 744--765.

\bibitem[ML06]{mcgraw2006}
R.~McGraw and Y.~Liu, \emph{Brownian drift-diffusion model for evolution of
  droplet size distributions in turbulent clouds}, Geophysical Research Letters
  \textbf{33} (2006), no.~3, L03802.

\bibitem[MS81]{matkowski1981}
B.~J. Matkowsky and Z.~Schuss, \emph{Eigenvalues of the fokker-planck operator
  and the approach to equilibrium for diffusions in potential fields}, SIAM
  Journal on Applied Mathematics \textbf{40} (1981), no.~2, 242--254.

\bibitem[NCCJ81]{north1981energy}
Gerald~R North, Robert~F Cahalan, and James~A Coakley~Jr, \emph{Energy balance
  climate models}, Reviews of Geophysics \textbf{19} (1981), 91--121.

\bibitem[Pav14]{pavliotisbook2014}
Grigorios~A. Pavliotis, \emph{{Stochastic Processes and Applications}},
  vol.~60, Springer, New York, 2014.

\bibitem[PS09]{paoli2009}
Roberto Paoli and Karim Shariff, \emph{Turbulent condensation of droplets:
  Direct simulation and a stochastic model}, Journal of the Atmospheric
  Sciences \textbf{66} (2009), no.~3, 723 -- 740.

\bibitem[PSK{\etalchar{+}}20]{prabhakaran2020}
Prasanth Prabhakaran, Abu Sayeed~Md Shawon, Gregory Kinney, Subin Thomas, Will
  Cantrell, and Raymond~A. Shaw, \emph{The role of turbulent fluctuations in
  aerosol activation and cloud formation}, Proceedings of the National Academy
  of Sciences \textbf{117} (2020), no.~29, 16831--16838.

\bibitem[Ris89]{risken}
Hannes Risken, \emph{{The Fokker-Planck Equation}}, second ed., Springer, 1989.

\bibitem[RY89]{rogers1989}
R.~R. Rogers and M.K. Yau, \emph{A short course in cloud physics}, 3rd ed. ed.,
  Pergamon Press Oxford, New York, 1989.

\bibitem[SBK17]{siewert_bec_krstulovic_2017}
Christoph Siewert, J{\'e}r{\'e}mie Bec, and Giorgio Krstulovic,
  \emph{Statistical steady state in turbulent droplet condensation}, Journal of
  Fluid Mechanics \textbf{810} (2017), 254--280.

\bibitem[Sha03]{shaw_2003}
R.A. Shaw, \emph{Particule-turbulence interactions in atmospheric clouds},
  Annual Review of Fluid Mechanics \textbf{35} (2003), no.~1, 183--227.

\bibitem[SPBC15]{sardina_2015}
Gaetano Sardina, Francesco Picano, Luca Brandt, and Rodrigo Caballero,
  \emph{Continuous growth of droplet size variance due to condensation in
  turbulent clouds}, Physical Review Letters \textbf{115} (2015), no.~18,
  184501.

\bibitem[Squ56]{squires_1956}
P.~Squires, \emph{The micro-structure of cumuli in maritime and continental
  air}, Tellus \textbf{8} (1956), no.~4, 443--444.

\bibitem[SS17]{siebert2017}
Holger Siebert and Raymond~A. Shaw, \emph{Supersaturation fluctuations during
  the early stage of cumulus formation}, Journal of the Atmospheric Sciences
  \textbf{74} (2017), no.~4, 975 -- 988.

\bibitem[VM09]{varnai2009modis}
Tam{\'a}s V{\'a}rnai and Alexander Marshak, \emph{{MODIS observations of
  enhanced clear sky reflectance near clouds}}, Geophysical Research Letters
  \textbf{36} (2009), no.~6, L06807.

\bibitem[WCD{\etalchar{+}}19]{weijer2019stability}
W.~Weijer, W.~Cheng, S.S Drijfhout, A.V. Fedorov, A.~Hu, L.C. Jackson, W.~Liu,
  E.L. McDonagh, J.V. Mecking, and J.~Zhang, \emph{{Stability of the Atlantic
  Meridional Overturning Circulation: A review and synthesis}}, Journal of
  Geophysical Research: Oceans \textbf{124} (2019), no.~8, 5336--5375.

\bibitem[WH06]{wallace_hobs}
J~M Wallace and P~V Hobbs, \emph{{Atmospheric Science, an Introductory
  Survey}}, Elsevier, 2006.

\bibitem[YSS{\etalchar{+}}24]{yang_2024}
F.~Yang, H.~F. Sadi, R.~A. Shaw, F.~Hoffmann, P.~Hou, A.~Wang, and
  M.~Ovchinnikov, \emph{Microphysics regimes due to haze-cloud interactions:
  cloud oscillation and cloud collapse}, EGUsphere \textbf{2024} (2024), 1--28.

\bibitem[ZMM{\etalchar{+}}20]{zelinka2020}
Mark~D. Zelinka, Timothy~A. Myers, Daniel~T. McCoy, Stephen Po-Chedley,
  Peter~M. Caldwell, Paulo Ceppi, Stephen~A. Klein, and Karl~E. Taylor,
  \emph{Causes of higher climate sensitivity in {CMIP6} models}, Geophysical
  Research Letters \textbf{47} (2020), no.~1, e2019GL085782.

\end{thebibliography}

\newcommand{\etalchar}[1]{$^{#1}$}
\providecommand{\bysame}{\leavevmode\hbox to3em{\hrulefill}\thinspace}
\providecommand{\MR}{\relax\ifhmode\unskip\space\fi MR }
\providecommand{\MRhref}[2]{%
  \href{http://www.ams.org/mathscinet-getitem?mr=#1}{#2}
}
\providecommand{\href}[2]{#2}

\end{document}